\documentclass[longauth]{aa} 
\usepackage{graphicx}
\usepackage{txfonts}
\usepackage{hyperref}

\usepackage{mychem}

\newcommand{\emm}[1]{\ensuremath{#1}}   
\newcommand{\emr}[1]{\emm{\mathrm{#1}}} 

\newcommand{\tint}[1][]{\emm{\Delta t_\emr{#1}}}
\newcommand{\beam}[1][]{\emm{\theta_\emr{#1}}}

\newcommand{\Msun}{\emm{M_\odot}}
\newcommand{\Msunperyr}{\emm{M_\odot\,\emr{yr^{-1}}}}

\newcommand{\ngmc}{$637$}

\def\mod#1{#1}

\makeatletter
\DeclareRobustCommand{\ion}[2]{%
  \text{#1\,\check@mathfonts\fontsize\sf@size\z@\selectfont #2}%
}
\makeatother

\newcommand{\hi}{\ion{H}{I}}

\newcommand{\FigRGB}{%
\begin{figure*}
\centering %
\includegraphics[width=0.95\linewidth]{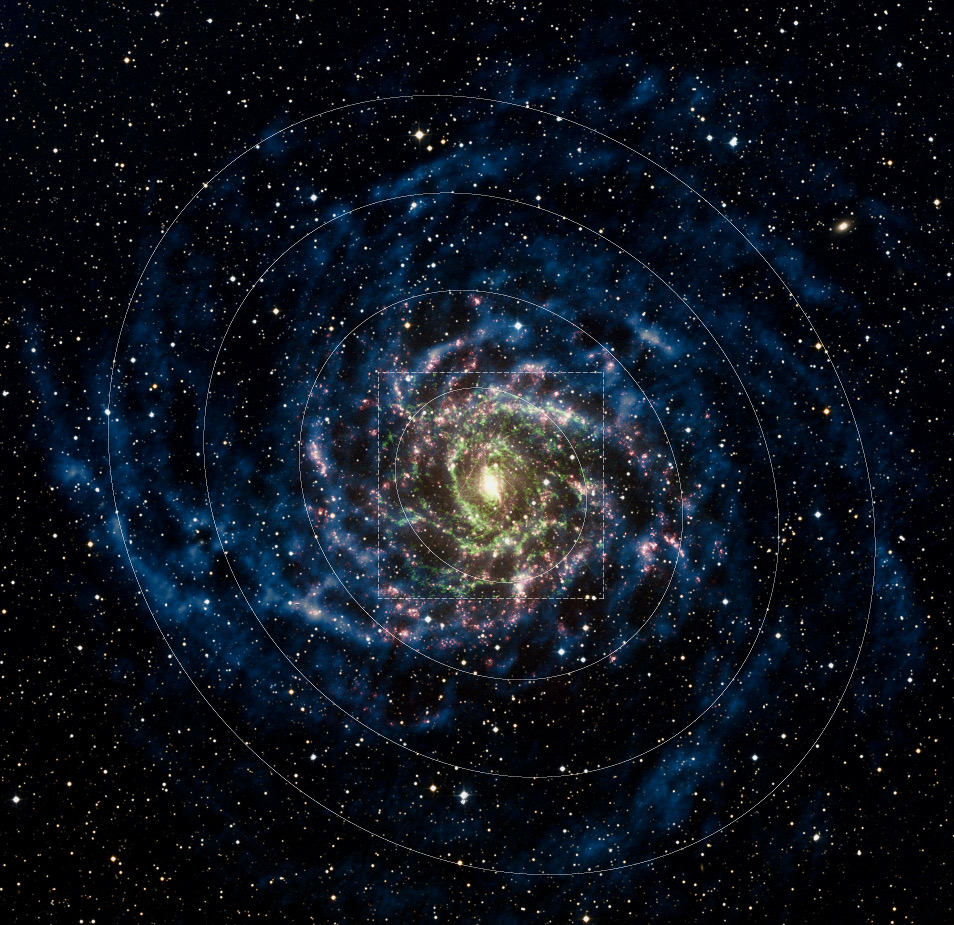}
\caption{Molecular, atomic, and ionized gas in IC\,342. This is a false-color composite of IC\,342 showing molecular gas traced by our new NOEMA \chem{CO}{10} survey (green), along with stellar light from DSS2 (white), atomic gas traced by the 21\,cm line using the VLA (blue), and H$\alpha$ emission from ionized gas (red). \mod{The dashed rectangle delimits} the field of view observed with NOEMA and solid ellipses show galactocentric radii at $5, 10, 15$, and $20$\,kpc.}
\label{fig:ic342:false-color}
\end{figure*}}

\newcommand{\FigData}{%
\begin{figure*}
\centering
\includegraphics[width=0.475\textwidth]{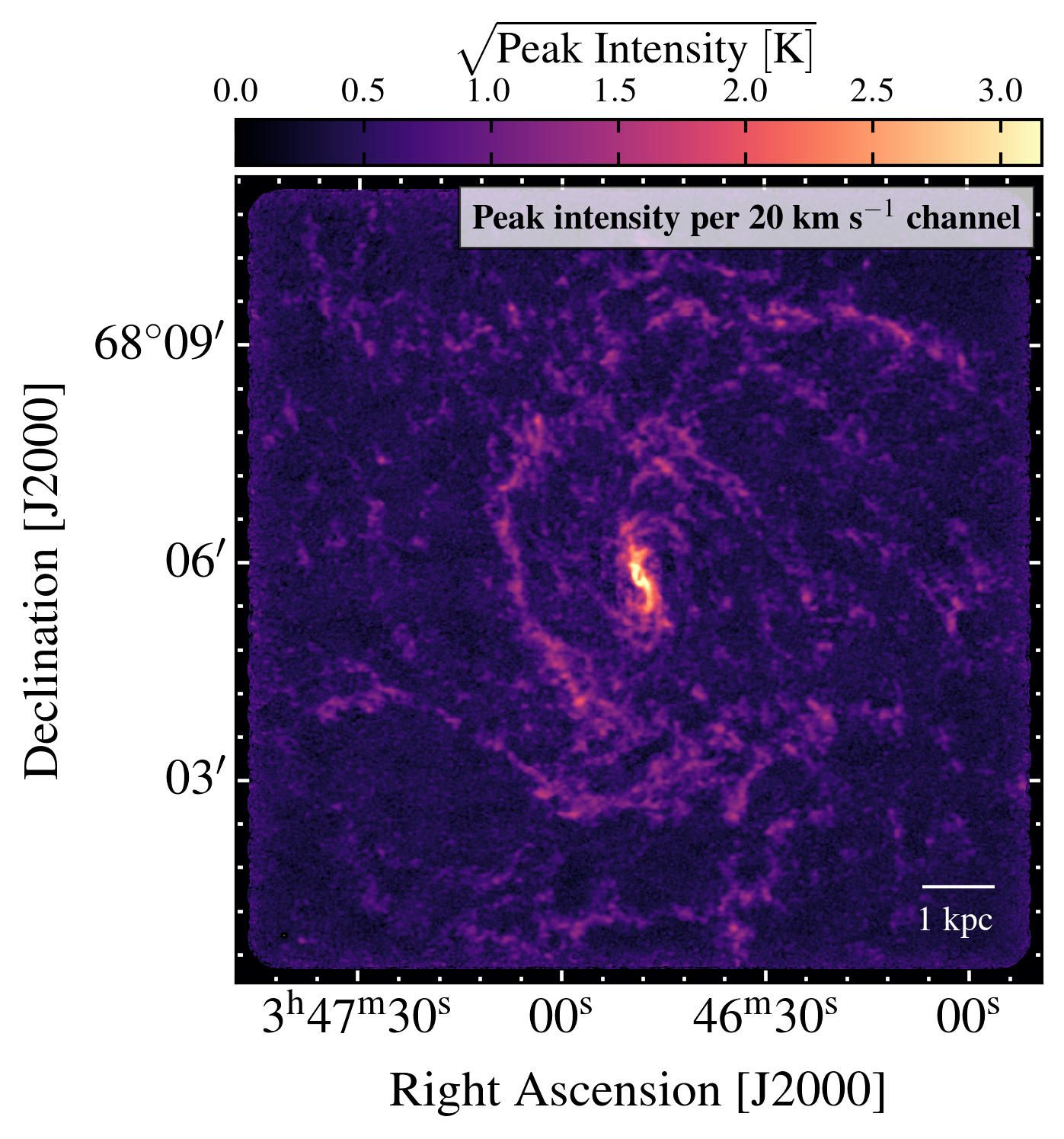}
\includegraphics[width=0.475\textwidth]{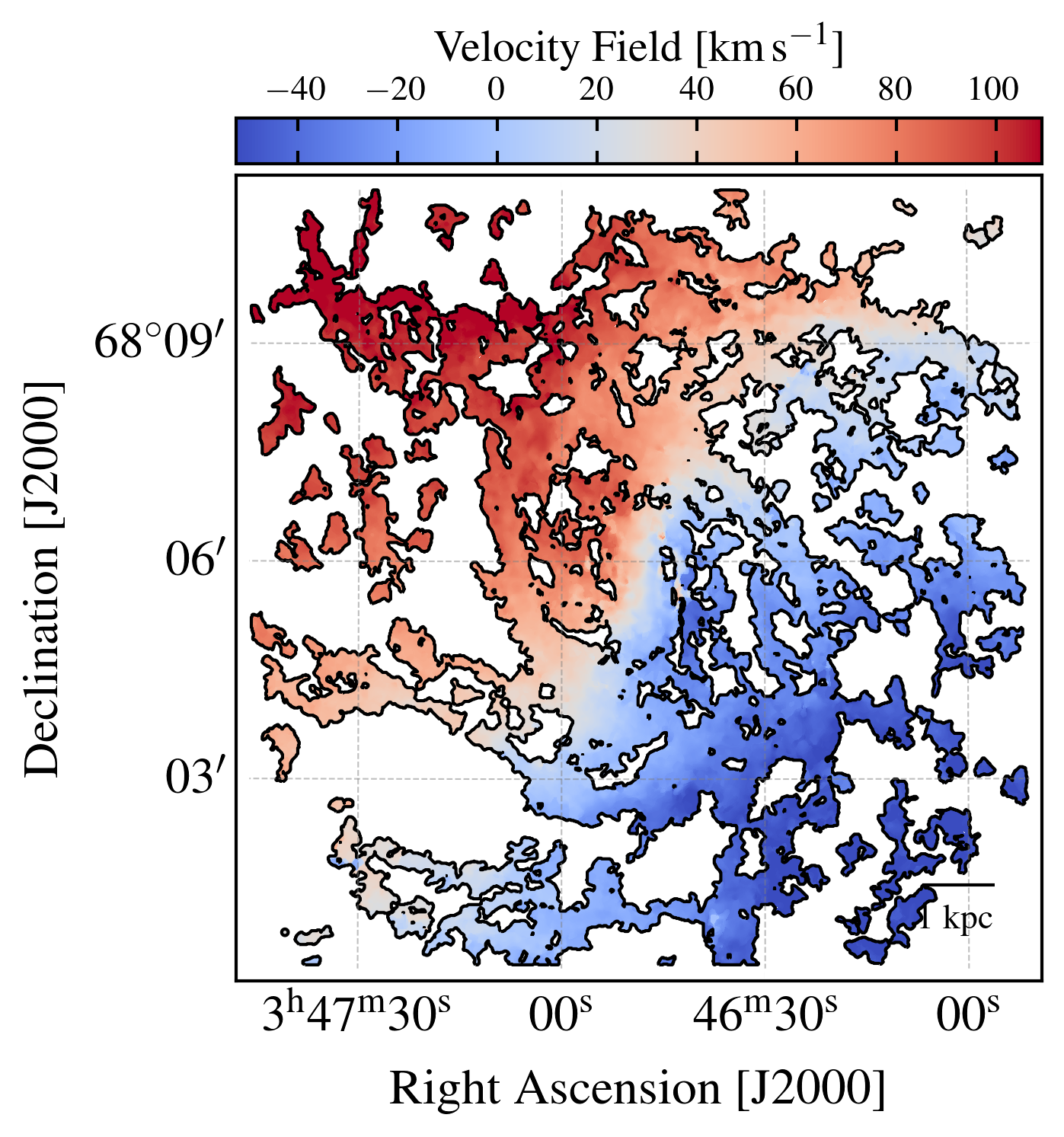} \\
\includegraphics[width=0.475\textwidth]{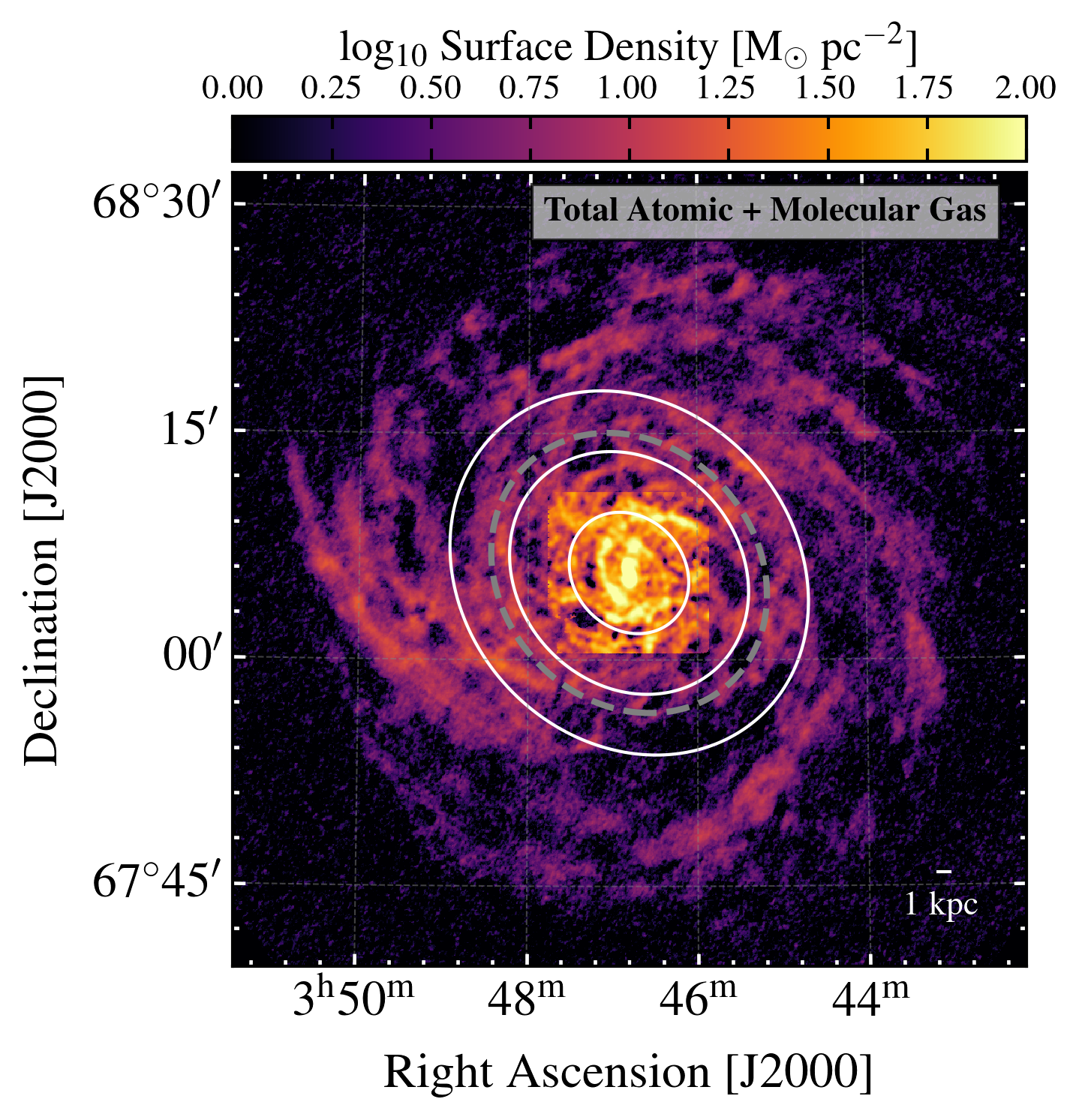}
\includegraphics[width=0.475\textwidth]{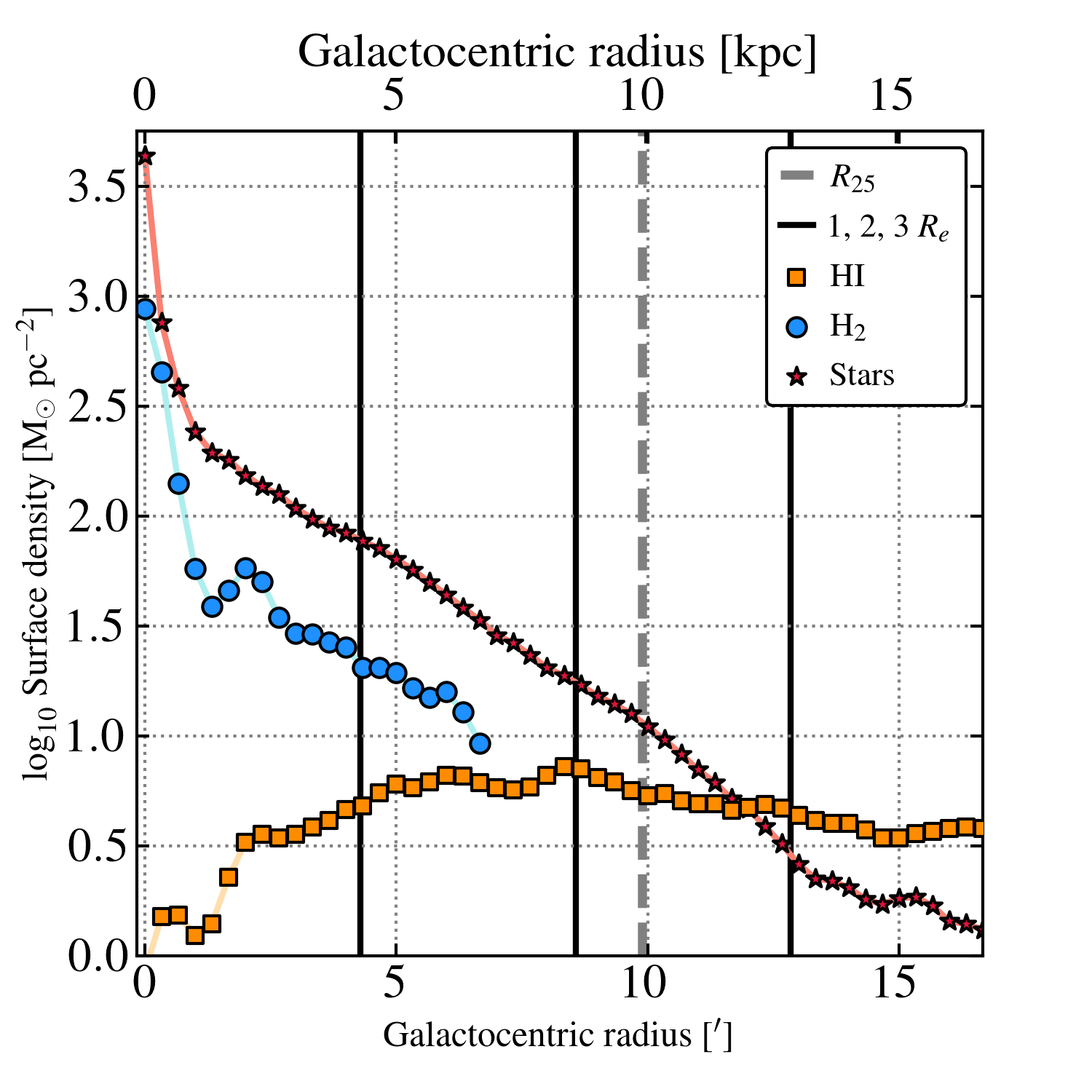}
\caption{{NOEMA survey of \chem{CO}{10} emission from IC\,342.} The distribution of the \chem{CO}{10} peak intensity at the native angular resolution (\textit{top left}) and the intensity-weighted mean CO velocity at $90$\,pc resolution (\textit{top right}) as revealed by our new CO survey. The images show spiral arms, abundant inter-arm emission, and a velocity field that mostly reflects a regularly rotating gas disk. The \textit{bottom} panels show the molecular gas traced by CO in the context of atomic gas and stars. In these panels, the CO and \hi{} data are shown at a common ${\sim}350$\,pc resolution. The \textit{bottom left} panel shows a map of the total neutral gas surface density adding molecular gas surface density to atomic gas surface density from \citet{CHIANG21}. The white ellipses show $1, 2, 3 \times R_e$ and the gray dashed ellipse indicates $1\,R_{25}$. The \textit{lower right} panel shows the azimuthally averaged mass surface density profiles for atomic gas, molecular gas, and stellar mass estimated from the near-infrared. The stellar mass distribution using \textit{HST} imaging is consistent with a nuclear star cluster and an exponential disk \citep{CARSON15}, while the presented profile suffers from the resolution of the \textit{WISE} data.  Our NOEMA survey covers the inner molecule-dominated region where the \hi{} emission is depressed, including the CO-bright center.
}
\label{fig:bigpicture}
\end{figure*}}

\newcommand{\FigGMCs}{%
\begin{figure*}[t!]
\centering
\includegraphics[width=0.43\textwidth]{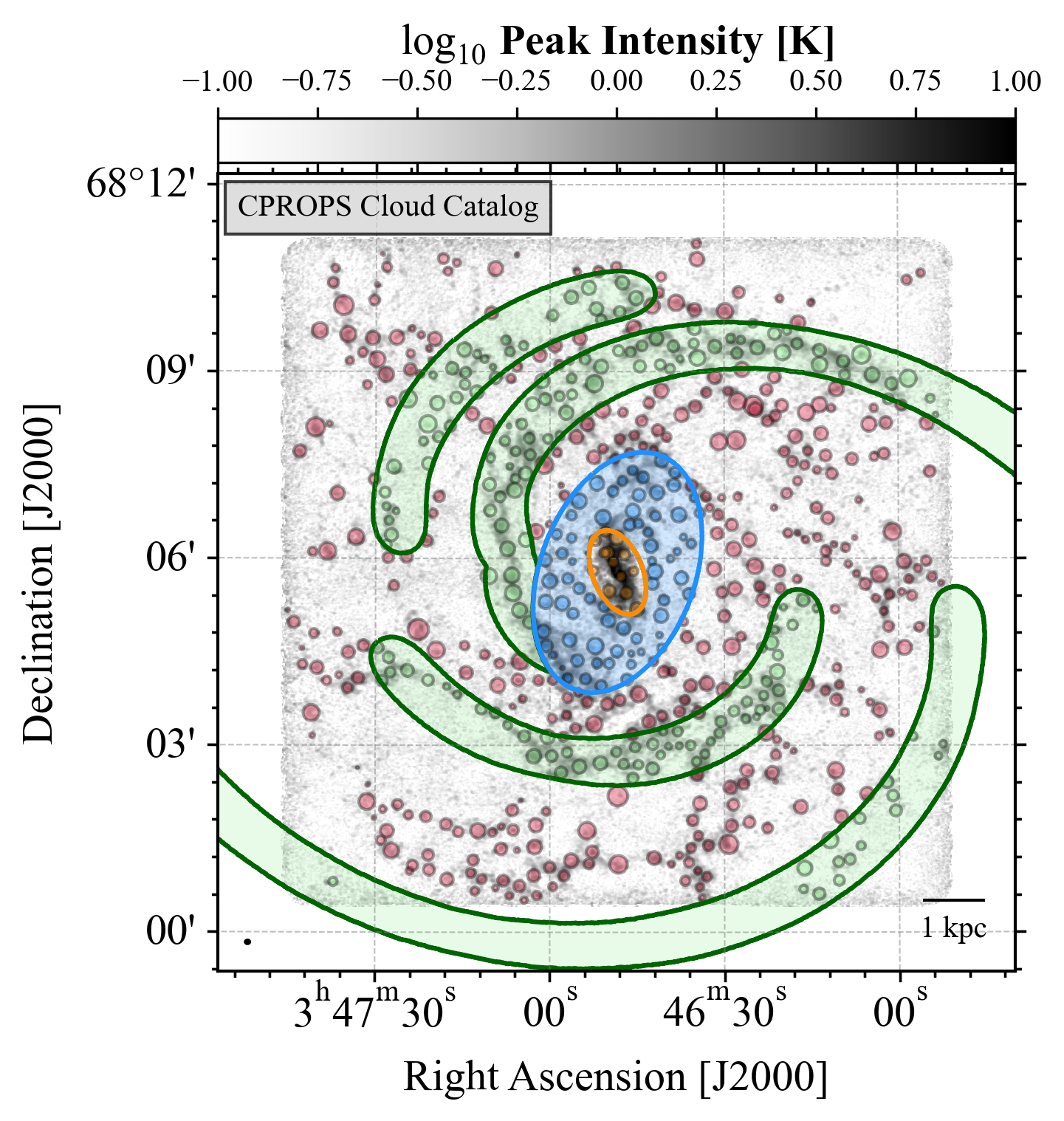}
\includegraphics[width=0.43\textwidth]{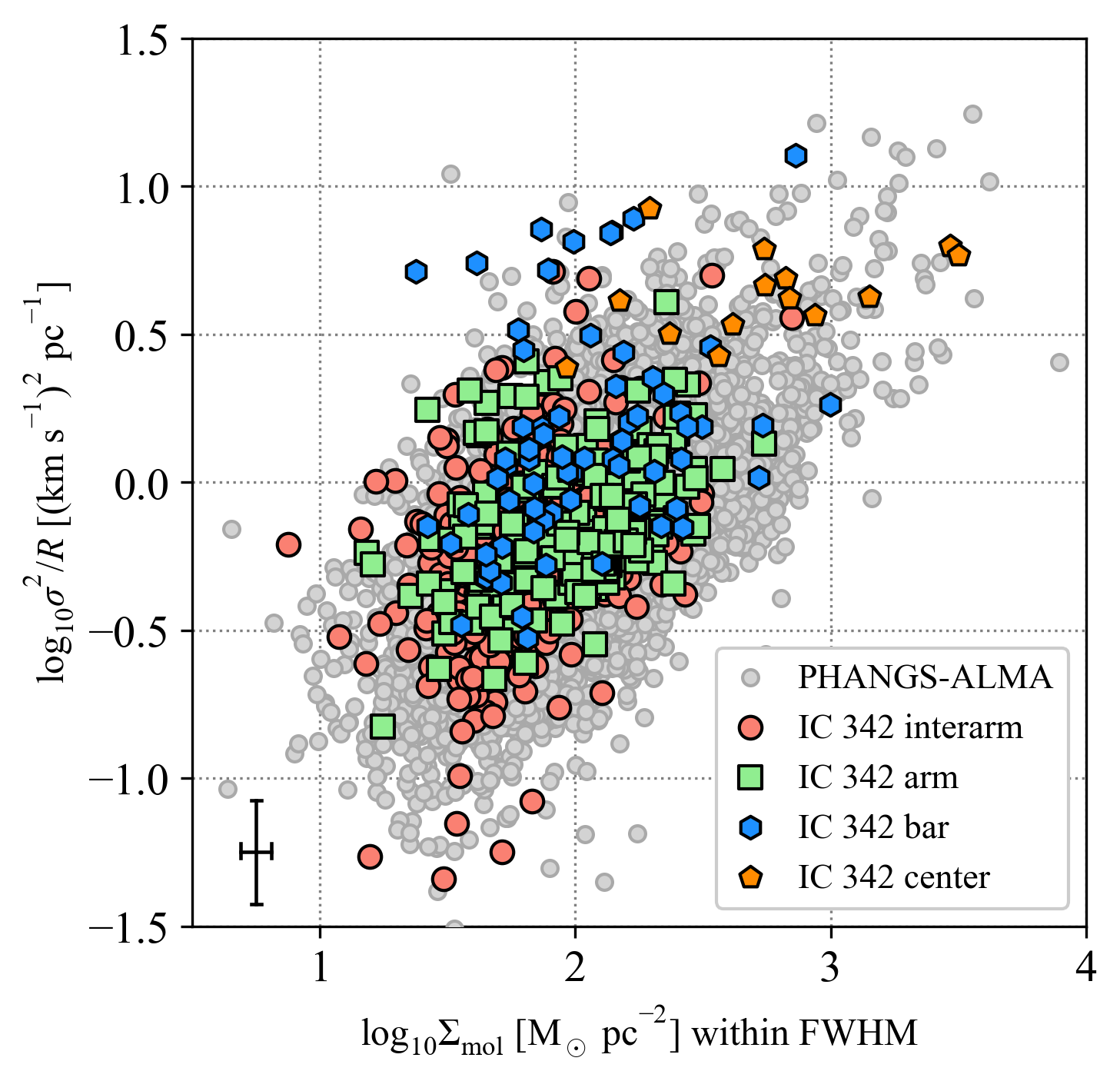} \\
\includegraphics[width=0.43\textwidth]{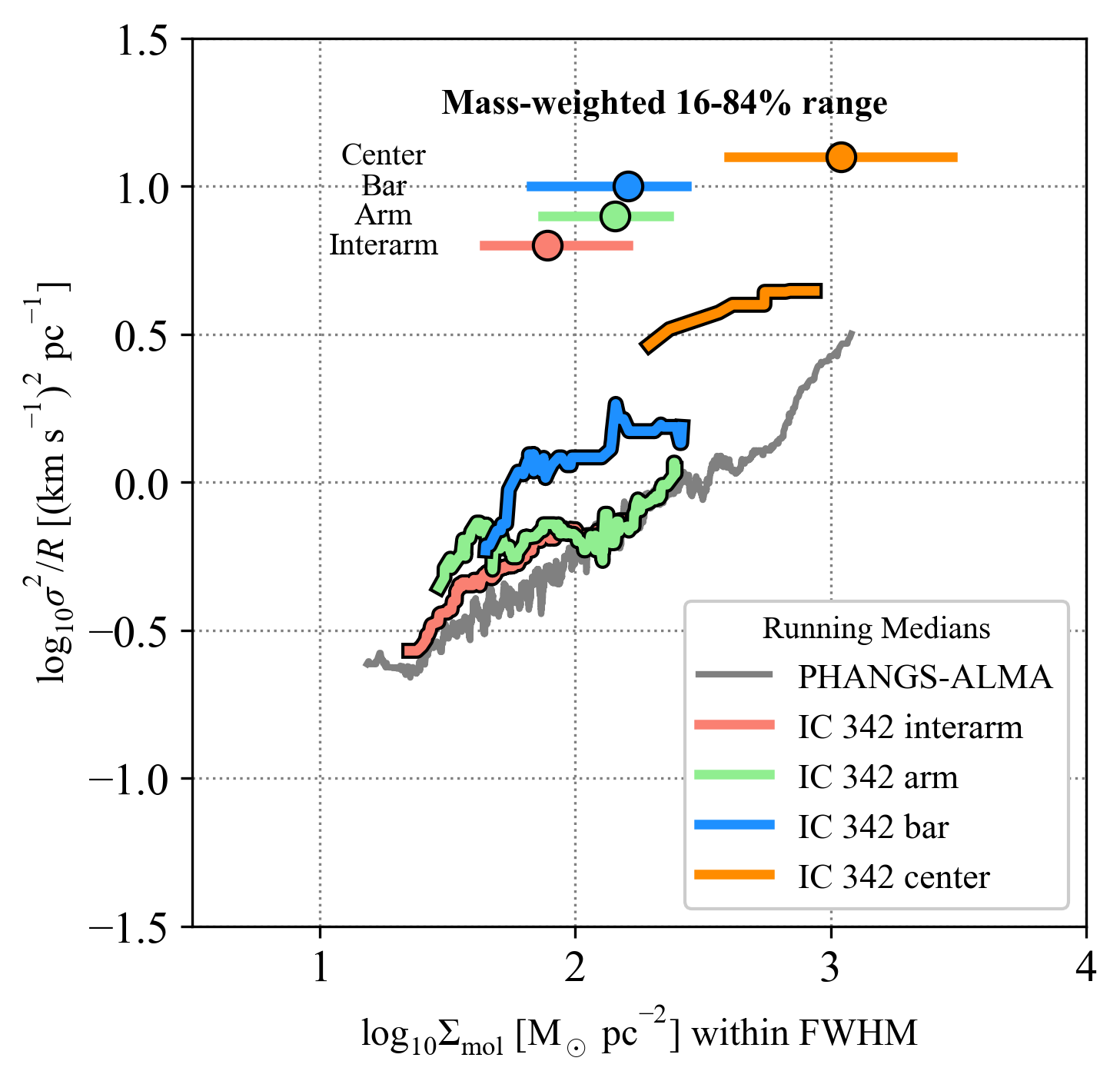}
\includegraphics[width=0.43\textwidth]{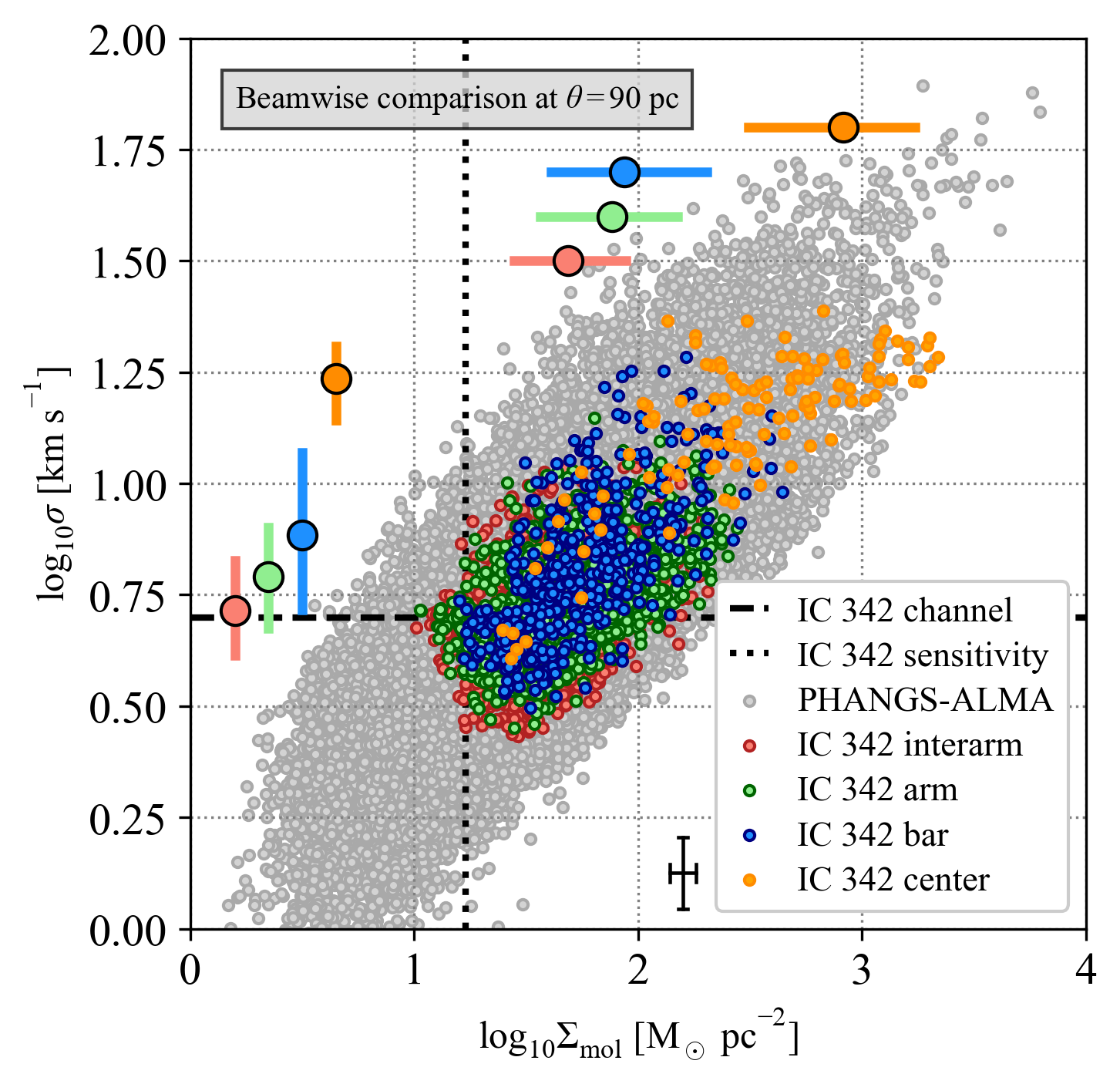}
\caption{{Giant molecular clouds or associations in IC\,342.} The top left panel shows the location of the molecular clouds that we catalog at $90$\,pc resolution. The size of each circle corresponds to the deconvolved radius and the color indicates which dynamical environment the cloud is assigned to (red shows interarm clouds, green arm cloud, and blue center clouds). The top right and bottom left panels show the cloud properties in $\sigma^2/R$ vs.\ $\Sigma_{\rm mol}$ space, in which clouds with a fixed virial parameter follow a diagonal line with a slope of unity in log--log space. \mod{$\Sigma_{\rm mol}$ is the average surface density within the FWHM size of each cloud, $\Sigma_{\rm mol} = M_{\rm CO} / (2\pi R^2)$.} We show the IC\,342 clouds for each environment and clouds at the same resolution and sensitivity, but better velocity resolution, from PHANGS--ALMA \citep[][and A. Hughes et al.\ in preparation]{ROSOLOWSKY21}. The bottom left panel replaces individual clouds with running medians. The bottom right panel adopts a beam-wise approach in which each independent line of sight at a fixed $90$\,pc resolution supplies a measurement of line width and surface density \citep{SUN18,SUN20B}. Again we compare the IC\,342 points to those from a large sample of PHANGS--ALMA galaxies and here we mark the channel width and approximate sensitivity limit for the IC\,342 data with dashed lines \mod{(for a Gaussian CO line, $\mathrm{FWHM}=2.35 \sigma$, so we always have more than one channel across the FWHM of the emission line)}. The bottom row also indicates the mass-weighted median \mod{(position of the color circles)} and $16{-}84$\% range \mod{(span of the color horizontal or vertical line)} for surface densities and line widths. \mod{These refer to the corresponding horizontal or vertical axis, with an arbitrary positioning along the perpendicular direction.} The \textit{mass-weighted} averages tend to lie at higher values than the bulk of the individual measurements because much of the molecular gas mass resides in a few high-mass clouds or associations. All panels tell a consistent story: arm clouds show mildly enhanced surface density and line width compared to inter-arm clouds, and the center shows significant enhancements in surface density and line width. The elevated line widths in the center indicate high virial parameters suggesting clouds with additional contributions to their line widths. The deviations from self-gravity virialization would be even more extreme if we adopted a CO-to-H$_2$ conversion factor below the Galactic value used to construct these plots. \mod{The black crosses in the right panels show representative error bars for our measurements in IC\,342, as explained in the main text.}}
\label{fig:clouds}
\end{figure*}}

\newcommand{\FigSFE}{%
\begin{figure}
\centering
\includegraphics[width=0.475\textwidth]
{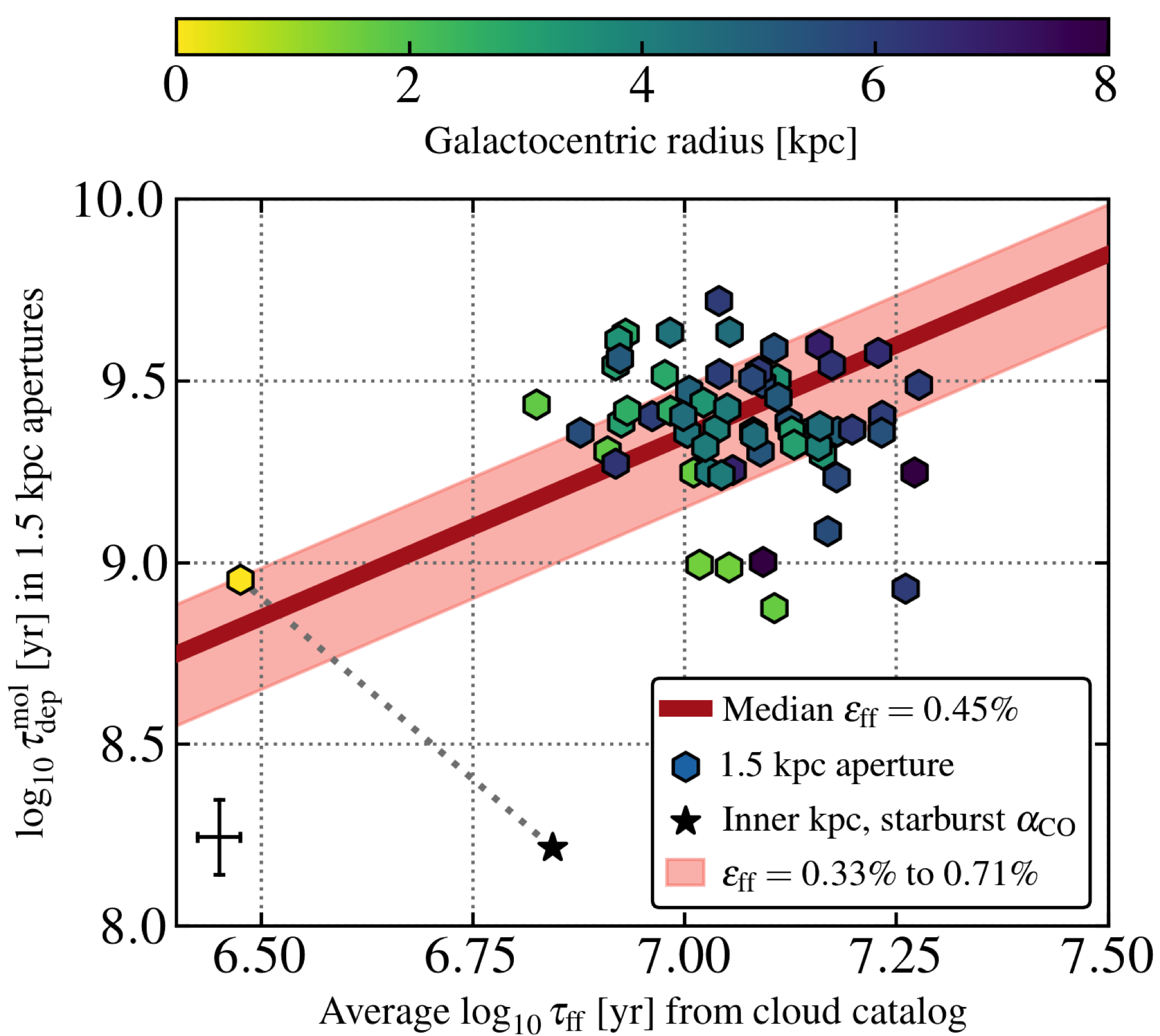}
\caption{{Star formation efficiency per gravitational free-fall time in $1.5$\,kpc regions.} Each hexagonal point shows the integrated molecular gas depletion time for a $1.5$\,kpc diameter region as a function of the mass-weighted average $\tau_{\rm ff}$ (see Eq.~\ref{eq:tau_ff}) for $90$\,pc resolution clouds in that region. A straight, diagonal line in this space, such as the dark red one, corresponds to a fixed $\epsilon_{\rm ff}$, which could be expected if density variations represent the primary drivers of depletion time variations. We color the regions by galactocentric radius and calculate the implied $\epsilon_{\rm ff}$ for each region. We estimate a median $\epsilon_{\rm ff}$ of $0.45\%$ with a $16{-}84$\% range of $0.33{-}0.71\%$, and we illustrate these with the solid red line and shaded pink region. For the central $1.5$\,kpc region, we illustrate the effect of switching from our adopted Galactic $\alpha_{\rm CO}$ to a starburst conversion factor. \mod{The black cross shows a representative error bar as explained in the main text. The plotted values can be found in Table~\ref{table:hexagons}.}}
\label{fig:eff}
\end{figure}}

\newcommand{\FigZoom}{%
\begin{figure*}
\centering
\includegraphics[width=0.95\linewidth]{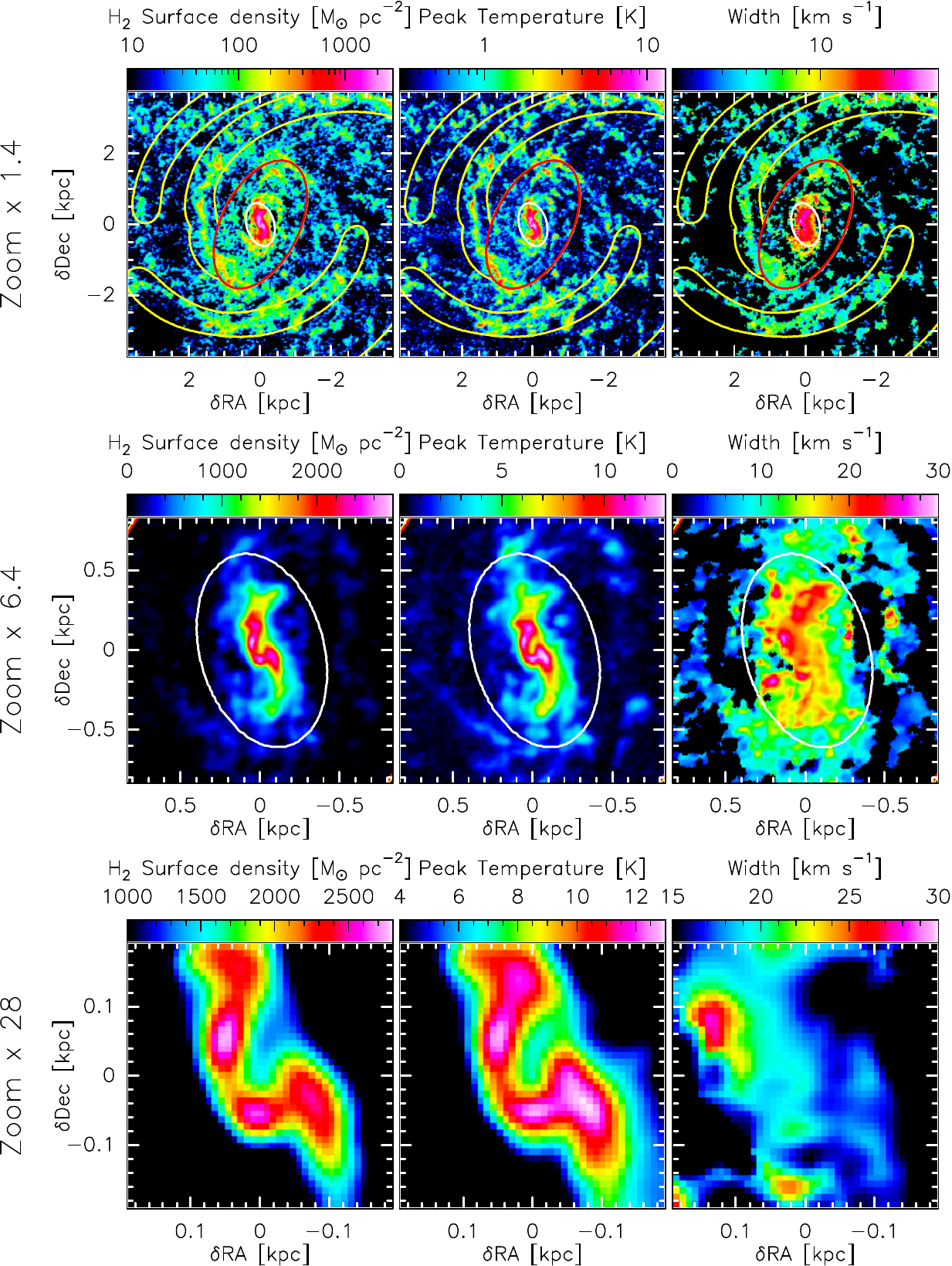}
\caption{Zooms of the molecular surface density, peak temperature, and line width towards the galaxy center. The images have been deprojected from the galaxy inclination on the plane of sky, and then converted to kpc using a distance of $3.45$\,Mpc. \mod{The yellow lines show our spiral mask, while the red and white ellipses delimit the extent of the bar and center environment, respectively (see Appendix~\ref{app:environment} for the definition of these environment)}.}
\label{fig:zoom}
\end{figure*}}

\newcommand{\FigNIR}{%
\begin{figure}
\centering
\includegraphics[width=\linewidth]{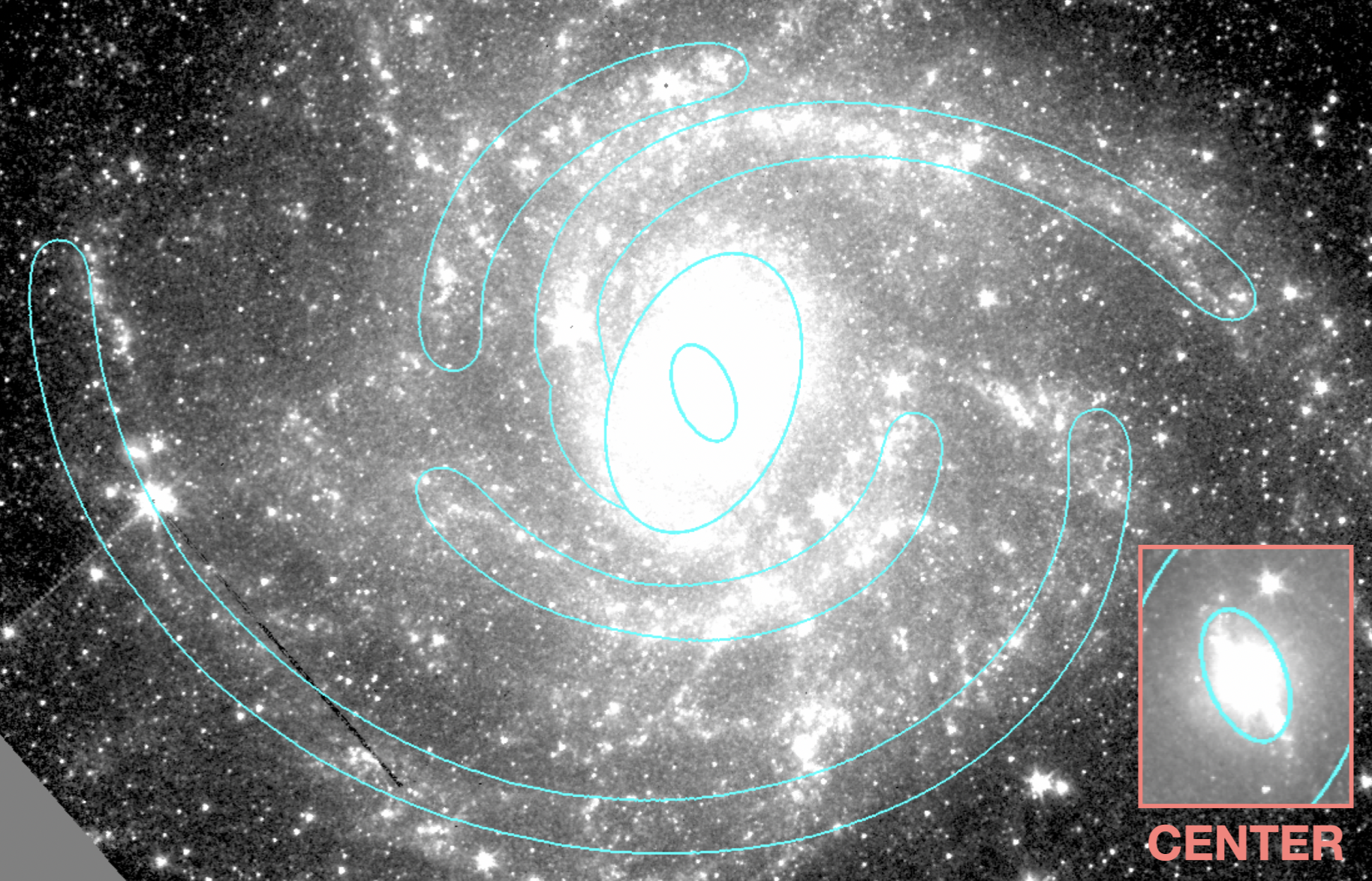}
\caption{\mod{\textit{Spitzer} IRAC 3.6\,$\mu$m image showing the definition of our adopted environments. The inset on the bottom-right highlights the center with a different color stretch.}
}
\label{fig:NIR}
\end{figure}}

\begin{document} 

\title{A sensitive, high-resolution, wide-field IRAM NOEMA CO(1-0) survey of the very nearby spiral galaxy IC\,342}
\titlerunning{NOEMA \chem{CO}{10} map of IC\,342}


\newcommand{\OSU}{\label{OSU} Department of Astronomy, The Ohio State University, 140 West 18th Avenue, Columbus, Ohio 43210, USA}

\newcommand{\Alberta}{\label{Alberta} Department of Physics, University of Alberta, Edmonton, AB T6G 2E1, Canada}

\newcommand{\ANU}{\label{ANU} Research School of Astronomy and Astrophysics, Australian National University, Canberra, ACT 2611, Australia}

\newcommand{\IPAC}{\label{IPAC} Caltech-IPAC, 1200 E. California Blvd. Pasadena, CA 91125, USA}

\newcommand{\Carnegie}{\label{Carnegi} Observatories of the Carnegie Institution for Science, 813 Santa Barbara Street, Pasadena, CA 91101, USA}

\newcommand{\CCAPP}{\label{CCAPP} Center for Cosmology and Astroparticle Physics, 191 West Woodruff Avenue, Columbus, OH 43210, USA}

\newcommand{\CfA}{\label{CfA} Harvard-Smithsonian Center for Astrophysics, 60 Garden Street, Cambridge, MA 02138, USA}

\newcommand{\CITEVA}{\label{CITEVA} Centro de Astronomía (CITEVA), Universidad de Antofagasta, Avenida Angamos 601, Antofagasta, Chile}

\newcommand{\CNRS}{\label{CNRS} CNRS, IRAP, 9 Av. du Colonel Roche, BP 44346, F-31028 Toulouse cedex 4, France}

\newcommand{\ESO}{\label{ESO} European Southern Observatory, Karl-Schwarzschild Stra{\ss}e 2, D-85748 Garching bei M\"{u}nchen, Germany}

\newcommand{\Heidelberg}{\label{Heidelberg} Astronomisches Rechen-Institut, Zentrum f\"{u}r Astronomie der Universit\"{a}t Heidelberg, M\"{o}nchhofstra\ss e 12-14, D-69120 Heidelberg, Germany}

\newcommand{\COOL}{\label{COOL} Cosmic Origins Of Life (COOL) Research DAO, coolresearch.io}

\newcommand{\ICRAR}{\label{ICRAR} International Centre for Radio Astronomy Research, University of Western Australia, 35 Stirling Highway, Crawley, WA 6009, Australia}

\newcommand{\IRAM}{\label{IRAM} Institut de Radioastronomie Millim\'{e}trique (IRAM), 300 Rue de la Piscine, F-38406 Saint Martin d'H\`{e}res, France}

\newcommand{\ITA}{\label{ITA} Universit\"{a}t Heidelberg, Zentrum f\"{u}r Astronomie, Institut f\"{u}r Theoretische Astrophysik, Albert-Ueberle-Str 2, D-69120 Heidelberg, Germany}

\newcommand{\IWR}{\label{IWR} Universit\"{a}t Heidelberg, Interdisziplin\"{a}res Zentrum f\"{u}r Wissenschaftliches Rechnen, Im Neuenheimer Feld 205, D-69120 Heidelberg, Germany}

\newcommand{\JHU}{\label{JHU} Department of Physics and Astronomy, The Johns Hopkins University, Baltimore, MD 21218, USA}

\newcommand{\Leiden}{\label{Leiden} Leiden Observatory, Leiden University, P.O. Box 9513, 2300 RA Leiden, The Netherlands}

\newcommand{\Maryland}{\label{Maryland} Department of Astronomy, University of Maryland, College Park, MD 20742, USA}

\newcommand{\MPE}{\label{MPE} Max-Planck-Institut f\"{u}r extraterrestrische Physik, Giessenbachstra{\ss}e 1, D-85748 Garching, Germany}

\newcommand{\MPIA}{\label{MPIA} Max-Planck-Institut f\"{u}r Astronomie, K\"{o}nigstuhl 17, D-69117, Heidelberg, Germany}

\newcommand{\Nagoya}{\label{Nagoya} Department of Physics, Nagoya University, Furo-cho, Chikusa-ku, Nagoya, Aichi 464-8602, Japan}

\newcommand{\NRAO}{\label{NRAO} National Radio Astronomy Observatory, 520 Edgemont Road, Charlottesville, VA 22903-2475, USA}

\newcommand{\OAN}{\label{OAN} Observatorio Astron\'{o}mico Nacional (IGN), C/Alfonso XII, 3, E-28014 Madrid, Spain}

\newcommand{\ObsParis}{\label{ObsParis} Sorbonne Universit\'{e}, Observatoire de Paris, Universit\'{e} PSL, CNRS, LERMA, F-75014, Paris, France}

\newcommand{\Princeton}{\label{Princeton} Department of Astrophysical Sciences, Princeton University, Princeton, NJ 08544 USA}

\newcommand{\UToledo}{\label{UToledo} University of Toledo, 2801 W. Bancroft St., Mail Stop 111, Toledo, OH, 43606}

\newcommand{\Toulouse}{\label{Toulouse} Universit\'{e} de Toulouse, UPS-OMP, IRAP, F-31028 Toulouse cedex 4, France}

\newcommand{\UBonn}{\label{UBonn} Argelander-Institut f\"ur Astronomie, Universit\"at Bonn, Auf dem H\"ugel 71, 53121 Bonn, Germany}

\newcommand{\UChile}{\label{UChile} Departamento de Astronom\'{i}a, Universidad de Chile, Camino del Observatorio 1515, Las Condes, Santiago, Chile}

\newcommand{\UConn}{\label{UConn} Department of Physics, University of Connecticut, Storrs, CT, 06269, USA}

\newcommand{\UCSD}{\label{UCSD} Center for Astrophysics and Space Sciences, Department of Physics,  University of California, San Diego, 9500 Gilman Drive, La Jolla, CA 92093, USA}

\newcommand{\UCSDAA}{\label{UCSDAA} Department of Astronomy \& Astrophysics,  University of California, San Diego, 9500 Gilman Drive, La Jolla, CA 92093, USA}

\newcommand{\UGent}{\label{UGent} Sterrenkundig Observatorium, Universiteit Gent, Krijgslaan 281 S9, B-9000 Gent, Belgium}

\newcommand{\ULyon}{\label{ULyon} Univ Lyon, Univ Lyon 1, ENS de Lyon, CNRS, Centre de Recherche Astrophysique de Lyon UMR5574,\\ F-69230 Saint-Genis-Laval, France}

\newcommand{\UMass}{\label{UMass} University of Massachusetts—Amherst, 710 N. Pleasant Street, Amherst, MA 01003, USA}

\newcommand{\UWyoming}{\label{UWyoming} Department of Physics and Astronomy, University of Wyoming, Laramie, WY 82071, USA}

\newcommand{\LAM}{\label{LAM} Aix Marseille Univ, CNRS, CNES, LAM (Laboratoire d’Astrophysique de Marseille), Marseille, France}

\newcommand{\UHawaii}{\label{UHawaii} Institute for Astronomy, University of Hawaii, 2680 Woodlawn Drive, Honolulu, HI 96822, USA}

\newcommand{\UCM}{\label{UCM} Departamento de F\'{\i}sica de la Tierra y Astrof\'{\i}sica, Universidad Complutense de Madrid, E-28040, Spain}

\newcommand{\IPARC}{\label{IPARC} Instituto de F\'{\i}sica de Part\'{\i}culas y del Cosmos IPARCOS, Facultad de Ciencias F\'{\i}sicas, Universidad Complutense de Madrid, E-28040, Spain}

\newcommand{\STScI}{\label{STScI} Space Telescope Science Institute, 3700 San Martin Drive, Baltimore, MD 21218, USA}

\newcommand{\McMaster}{\label{McMaster} Department of Physics and Astronomy, McMaster University, 1280 Main Street West, Hamilton, ON L8S 4M1, Canada}

\newcommand{\INAF}{\label{INAF} INAF -- Osservatorio Astrofisico di Arcetri, Largo E. Fermi 5, I-50157, Firenze, Italy}

\newcommand{\Sydney}{\label{Sydney} Sydney Institute for Astronomy, School of Physics A28, The University of Sydney, NSW 2006, Australia}

\newcommand{\UA}{\label{UA} Centro de Astronomía (CITEVA), Universidad de Antofagasta, Avenida Angamos 601, Antofagasta, Chile}

\newcommand{\CITA}{\label{CITA} Canadian Institute for Theoretical Astrophysics (CITA), University of Toronto, 60 St George St, Toronto, ON M5S 3H8, Canada}

\newcommand{\ASIAA}{\label{ASIAA} Institute of Astronomy and Astrophysics, Academia Sinica, No. 1, Sec. 4, Roosevelt Road, Taipei 10617, Taiwan}

\newcommand{\TKU}{\label{TKU} Department of Physics, Tamkang University, No.151, Yingzhuan Rd., Tamsui Dist., New Taipei City 251301, Taiwan}

\newcommand{\PSMA}{\label{PSMA} Penn State Mont Alto, 1 Campus Drive, Mont Alto, PA  17237, USA}

\newcommand{\ILL}{\label{ILL} Institut Laue-Langevin, 71 avenue des Martyrs, F-38042 Grenoble, France}

\newcommand{\TUM}{\label{TUM} Technical University of Munich, School of Engineering and Design, Department of Aerospace and Geodesy, Chair of Remote Sensing Technology, Arcisstr. 21, 80333 Munich, Germany}


\author{%
Miguel~Querejeta\inst{\ref{OAN}}              
\and Jérôme~Pety\inst{\ref{IRAM},\ref{ObsParis}}    
%
\and Andreas~Schruba\inst{\ref{MPE}}
\and Adam~K.~Leroy\inst{\ref{OSU}}            
\and Cinthya~N.~Herrera\inst{\ref{ILL},\ref{IRAM}}  
\and I-Da~Chiang\inst{\ref{ASIAA},\ref{UCSD}} 
\and Sharon~E.~Meidt\inst{\ref{UGent}}        
\and Erik~Rosolowsky\inst{\ref{Alberta}}      
\and Eva~Schinnerer\inst{\ref{MPIA}}          %
\and Karl~Schuster\inst{\ref{IRAM}}           %
\and Jiayi~Sun\inst{\ref{McMaster},\ref{CITA}}  
\and Kimberly~A.~Herrmann\inst{\ref{PSMA}}
%
%
\and Ashley~T.~Barnes\inst{\ref{UBonn}}      
\and Ivana~Be\v{s}li\'c\inst{\ref{UBonn}}     
\and Frank~Bigiel\inst{\ref{UBonn}}           
\and Yixian~Cao\inst{\ref{MPE}}               
\and M\'elanie~Chevance\inst{\ref{ITA},\ref{COOL}}
\and Cosima~Eibensteiner\inst{\ref{UBonn}}    
\and Eric~Emsellem\inst{\ref{ESO},\ref{ULyon}} %
\and Christopher~M.~Faesi\inst{\ref{UConn}}   
\and Annie~Hughes\inst{\ref{CNRS},\ref{Toulouse}}
\and Jaeyeon~Kim\inst{\ref{ITA}}       
\and Ralf~S.~Klessen\inst{\ref{ITA},\ref{IWR}}
\and Kathryn~Kreckel\inst{\ref{Heidelberg}}   
\and J.~M.~Diederik~Kruijssen\inst{\ref{TUM}, \ref{COOL}} 
\and Daizhong~Liu\inst{\ref{MPIA}}            %
\and Nadine~Neumayer\inst{\ref{MPIA}}         
\and Hsi-An~Pan\inst{\ref{TKU}}               
\and Toshiki~Saito\inst{\ref{MPIA}}           
\and Karin~Sandstrom\inst{\ref{UCSDAA}}         
\and Yu-Hsuan~Teng\inst{\ref{UCSD}}           
\and Antonio~Usero\inst{\ref{OAN}}            
\and Thomas~G.~Williams\inst{\ref{MPIA}}      
\and Antoine~Zakardjian\inst{\ref{Toulouse}}  %
}

\institute{\OAN{} \and \IRAM{} \and \ObsParis{} \and \MPE{} \and \OSU{} \and \ILL{}
\and \ASIAA{}  \and \UCSD{} \and \UGent{} \and \Alberta{} \and
\MPIA{} \and \McMaster{} \and \CITA{} \and \PSMA{} \and \UBonn{} \and \ITA{} \and
\COOL{} \and \ESO{} \and
\ULyon{} \and \UConn{} \and \CNRS{} \and \Toulouse{} \and \IWR{} \and \Heidelberg{} \and \TUM{} \and \TKU{}  \and \UCSDAA{} }


\abstract{
We present a new wide-field $10.75\times10.75$\,arcmin$^2$ ($\approx 11\times11$\,kpc$^2$), high-resolution (\mod{$\theta = 3.6\arcsec \approx 60$\,pc}) NOEMA \chem{CO}{10} survey of the very nearby ($d=3.45$\,Mpc) spiral galaxy IC\,342. The survey spans out to about $1.5$ effective radii and covers most of the region where molecular gas dominates the cold interstellar medium. We resolved the CO emission into ${>}600$ individual giant molecular clouds and associations. We assessed their properties and found that overall the clouds show approximate virial balance, with typical virial parameters of $\alpha_{\rm vir} = 1{-}2$. The typical surface density and line width of molecular gas increase from the inter-arm region to the arm \mod{and bar} region, and they reach their highest values in the inner 
\mod{kiloparsec} of the galaxy {(median $\Sigma_{\rm mol} \approx 80, 140, 160$, and $1100\,M_\odot$\,pc$^{-2}$, $\sigma_{\rm CO} \approx 6.6$, $7.6$, $9.7$, and $18.4$\,km\,s$^{-1}$ for inter-arm, arm, bar, and center clouds, respectively)}. Clouds in the central part of the galaxy show an enhanced line width relative to their surface densities and evidence of additional sources of dynamical broadening. All of these results agree well with studies of clouds in more distant galaxies at a similar physical resolution. Leveraging our measurements to estimate the density and gravitational free-fall time at $90$\,pc resolution,
{averaged on 1.5\,kpc hexagonal apertures,}
we estimate a typical star formation efficiency per free-fall time of $0.45\%$ with a $16{-}84$\% variation of $0.33{-}0.71\%$ 
{among such $1.5$\,kpc regions.}
\mod{We speculate that bar-driven gas inflow could explain the large gas concentration in the central kiloparsec and the buildup of the massive nuclear star cluster.}
This wide-area CO map of the closest face-on massive spiral galaxy demonstrates the current mapping power of NOEMA and has many potential applications. The data and products are publicly available.
}

\keywords{Galaxies -- Giant Molecular Clouds -- galaxies: ISM -- galaxies: star formation}

\maketitle


\section{Introduction}

\begin{table}
  \centering %
  \caption{Global properties of IC\,342.}
  \resizebox{\linewidth}{!}{%
    \begin{tabular}{ccc}
      \hline
      \hline
      Property                                                & Value                                    & Reference \\
      \hline
      Hubble Type                                             & SABcd                                    & NED/LEDA \\
      Distance                                                & $3.45 \pm 0.13$\,Mpc                     & \citet{ANAND21} \\
      Scale                                                   & $1\arcsec \sim 16.7$\,pc                 & for above distance\\
      Center R.\,A.                                           & $03^\emr{h}\,46^\emr{m}\,48.50^\emr{s}$  & NED \\
      Center Dec.                                             & $+68\degr\,05\arcmin\,46.9\arcsec$                  & NED \\
      Systemic Vel.                                           & $30 \pm 2$\,km\,s$^{-1}$                 & \citet{MEIDT09} \\
      Systemic Vel.                                           & $32 \pm 2$\,km\,s$^{-1}$                 & this work \\
      Inclination                                             & $31 \pm 5\deg$                           & \citet{MEIDT09} \\
      Position Angle                                          & $42 \pm 3\deg$                           & \citet{MEIDT09} \\
      $R_{25}$                                                & $9.98\arcmin \pm 0.46\arcmin$                        & LEDA \\
      Effective Radius                                        & $4.28\arcmin \approx 4.3$\,kpc                                  & this work for above distance \\
      $E(B{-}V)_\emr{foreground}$                             & $0.494$\,mag                             & \citet{SCHLAFLY11} \\
      $E(B{-}V)_\emr{internal}$                               & $0.507$\,mag                             & this work \\
      $12+\emr{log(O/H)}_{R_0}$                               & $8.83 \pm 0.04$\,dex                     & \citet{PILYUGIN14} \\
      $\Delta \emr{log(O/H)}/R_\emr{gal}$\tablefootmark{a}    & $-0.513 \pm 0.087$\,dex/$R_{25}$         & \citet{PILYUGIN14} \\
      $M_\emr{star}$                                          & $1.8 \times 10^{10}$~\Msun               & this work \\
      $M_\emr{atom}$                                          & $8.5 \times 10^9$~\Msun                  & \citet{CROSTHWAITE00} \\
      $M_\emr{mol}$\tablefootmark{b}                          & $(4.2 \pm 0.8) \times 10^9$~\Msun        & this work \\
      $M_\emr{dust}$                                          & $(4.5 \pm 0.9) \times 10^7$ \Msun        & \citet{ANIANO20} \\
      SFR(22$\,\mu$m) or SFR(TIR)                             & $1.75 \pm 0.25$ \Msunperyr               & this work/\citet{SANDERS03} \\
      \hline
    \end{tabular}
  }
  \label{tab:ic342}
  \tablefoot{
Morphological Hubble type, distance, and kinematic orientation parameters of IC\,342 according to the reference provided, the NASA/IPAC Extragalactic Database (NED) \footnote{\url{https://ned.ipac.caltech.edu}}, or LEDA \footnote{\url{https://leda.univ-lyon1.fr}}. $R_{25}$ is the 25 mag/arcsec$^2$ $B$-band isophotal radius, while the effective radius contains half of the light. $E(B - V)$ represents optical extinction, $12 + \log({\rm O/H})_{R_0}$ is the gas-phase metallicity, and $\Delta \emr{log(O/H)}/R_\emr{gal}$ is the metallicity radial gradient (where $R_\emr{gal}$ is galactocentric radius). $M_\emr{star}$, $M_\emr{atom}$, $M_\emr{mol}$, and $M_\emr{dust}$ represent the total stellar, atomic, molecular, and dust mass in this galaxy. SFR is the total star formation rate.  All masses were scaled to our adopted distance.
$^{(a)}$ Oxygen abundance gradient constrained by only five~HII regions \citep{MCCALL85}. %
$^{(b)}$ We estimated the molecular gas mass using our short-spacing-corrected NOEMA data (total CO luminosity multiplied by the Galactic $\alpha_{\rm CO}$). Since the NOEMA field of view is limited, we extrapolated to the whole galaxy according to the proportion of the total WISE 22$\,\mu$m luminosity that falls within the NOEMA field of view; as WISE 22$\,\mu$m is a SFR tracer, it should roughly scale according to the molecular mass at each radius.
} %
\end{table}

Giant molecular clouds (GMCs) represent the immediate sites of star formation and set the boundary conditions for star formation \citep[e.g.,][]{KENNICUTT12}. 
{Far from universal, the properties of GMCs strongly depend on the local galactic environment \citep[e.g.,][]{GRATIER10,HUGHES13,COLOMBO14,ROSOLOWSKY21}.}
In theory, the properties of molecular clouds play a key role in setting the rate at which molecular gas can form stars \citep[e.g.,][]{PADOAN12,FEDERRATH13,KRUMHOLZ19}. These theoretical predictions have spurred a number of recent observational attempts to quantitatively link the properties of individual molecular clouds to the rate of star formation in galaxies or parts of galaxies \citep[e.g.,][]{BARNES17,LEROY17A,OCHSENDORF17,HIROTA18,KRECKEL18,UTOMO18,SCHRUBA19,PESSA21}.

To make such measurements, one must survey the properties of molecular gas with high resolution and high completeness across large parts of galaxies. Most often, this is done by observing low-$J$ CO line emission, the standard tracer of molecular gas in galaxies \citep[e.g.,][]{BOLATTO13B}. Though CO is among the brightest millimeter-wave lines, CO surveys still represent a major technical challenge, requiring a focused effort using the best millimeter and submillimeter telescopes in the world.

\FigRGB{} %
\FigData{} %

At a distance of only $d = 3.45$\,Mpc \citep{ANAND21}, IC\,342 represents one of the closest, massive, star-forming spiral galaxies beyond the Local Group (see Table~\ref{tab:ic342}). Among massive, vigorously star-forming disk galaxies, its proximity is rivalled only by Andromeda, NGC\,253, and NGC\,4945 and all of these systems are far more inclined than IC\,342. This proximity and face-on orientation ($i\sim30^\circ$) translate into a sharper physical resolution and less confusion due to galaxy projection across the entire electromagnetic spectrum. Even among nearby northern galaxies, IC\,342 stands out as it is a prime example of a bulge-less spiral galaxy hosting a massive, compact nuclear star cluster \citep{CARSON15,NEUMAYER20} which is among the brightest ones known in nearby late-type galaxies \citep{GEORGIEV14}. Indeed, were it not for its position behind the Galactic plane ($b=10.6^\circ$), IC\,342 would certainly be among the best-known galaxies in the sky. Even given this situation, the gas-rich nuclear region of IC\,342 has been a key target of millimeter-wave studies for decades {\citep[e.g.,][]{LO84,ECKART90,DOWNES92,SAKAMOTO99,MEIER05,PAN14}}. This rich region hosts one of the nearest nuclear star clusters that appears to heavily impact its surrounding gas reservoir \citep{SCHINNERER03} and the nuclear region of IC\,342 is one of the closest starbursts (we estimated a star formation rate inside the central 1.5\,kpc equal to ${\sim}0.2\,M_{\odot}\,\mathrm{yr}^{-1}$; see section~\ref{subsec:supporting}). 
{Thus, there are some similarities with the less massive, very nearby galaxies M33 and NGC\,5253 regarding the presence of nuclear stellar clusters and nuclear starbursts. In any case,}
the proximity, orientation, and active star formation that allowed these earlier studies make IC\,342 an ideal target for wider area CO surveys using more modern facilities.

Here, we report on a new NOrthern Extended Millimeter Array (NOEMA) wide-area survey of CO emission of IC\,342. We used NOEMA, supported by the IRAM~\mbox{30-m} telescope, to observe \chem{CO}{10} emission from ${\sim}1000$ individual pointings across the inner region of IC\,342. In total, we map $10.75\times10.75$\,arcmin$^2$, that is, about $11\times11$\,kpc$^2$, at an angular resolution of ${\sim}3.6\arcsec$ (or ${\sim}60$\,pc). Here we describe this new survey, present a catalog of \ngmc\ molecular clouds from the galaxy, and compare their properties across different environments. We also assess how well environmental variations in the mean surface density of molecular clouds predict apparent variations in the molecular gas depletion time, testing some of the theoretical predictions described above.

\mod{IRAM NOEMA is the most powerful millimeter-wave interferometer in the northern hemisphere and one of the few telescopes in the world with the ability to conduct such CO surveys. With twelve \mbox{15-m} dishes in the French Alps, recently upgraded receivers, and efficient observing, NOEMA can rapidly survey molecular line emission over hundreds or thousands of individual pointings. While the capabilities of ALMA are widely appreciated, NOEMA can survey CO emission at high completeness and high resolution from crucial northern targets such as the prototypical starburst M82, the iconic spirals M51 and M101, the Local Group galaxies M31 and M33, and the ``fireworks'' galaxy NGC\,6946 (all northern with $d < 10$\,Mpc). The Plateau de Bure Interferometer, NOEMA's predecessor, demonstrated these capabilities with the PAWS survey of M51 \citep{SCHINNERER13,PETY13} and a 2-3\,mm molecular line survey of NGC\,6946 \citep{EIBENSTEINER2022}. \citet{KRIEGER21} recently used NOEMA to produce a wide-field, high resolution map of CO emission in M82, and \citet{BESLIC21} mapped with NOEMA the $J=1{-}0$ lines of two CO isotopologues (\chem{^{13}CO} and \chem{C^{18}O}) and of high critical density molecular tracers (\chem{HCN}, \chem{HNC}, and \chem{HCO^+}) towards the nearby strongly barred galaxy NGC\,3627.}

\section{Data}

\subsection{\mod{Molecular gas surface density: The} NOEMA survey}

As part of project S16BF (P.I.\ A.~Schruba), we observed the late-type
spiral galaxy IC\,342 for 120\,hours using eight~antennas in the
D~configuration of NOEMA during the summer semester of 2016. 
A mosaic of 941 pointings was required to
correctly sample the $10.75 \times 10.75$\,arcmin$^2$ targeted field of
view. The short spacings that were filtered out by the interferometer were provided
from observations with the IRAM \mbox{30-m} telescope during 43\,hours in
July 2016.

The data processing (calibration, merging, imaging, and deconvolution) followed the standard procedures of both observatories. Details are given in Appendix~\ref{sec:obs+red}. 
{Briefly, data were calibrated using GILDAS/CLIC. We used reference quasars as phase and amplitude calibrators, and bright Galactic sources as flux calibrators. Around 20\% of the data were flagged due to unstable weather. Imaging was carried out using GILDAS/MAPPING. First, we constructed pseudo-visibilities out of the IRAM 30m data and they were merged with the NOEMA data in the $uv$ plane using the task \texttt{uvshort}. The imaging process involved cleaning with the H\"ogbom algorithm using a broad mask based on a smoothed and clipped version of the IRAM 30m cube. The resulting data cube has an rms of 0.11 K over 5\,km\,s$^{-1}$ channels.}
We constructed data cubes with 5, 10, and 20\,km\,s$^{-1}$ spectral resolution. These three cubes were deconvolved separately. Because the deconvolution becomes more accurate when the signal-to-noise ratio of the data increases, we were able to deconvolve more flux in our $20$\,km\,s$^{-1}$ resolution cube. The three cubes share the same angular resolution of $4\times3.25$\,arcsec$^2$ with a position angle of $91^\circ$. Table~\ref{tab:obs} lists the salient features of the observations and reports the properties of our 5\,km\,s$^{-1}$ resolution data cube.

\mod{We compared the flux between the NRO 45m data at $20''$-resolution
published in the Nobeyama CO Atlas of Nearby Spiral Galaxies \citep{KUNO07} with the IRAM 30m data at $22.5''$ resolution described here. We used the same channel spacing of 5\,km\,s$^{-1}$. We computed the flux in a field of view of $600''\times600''$ centered on the galaxy center and within the velocity range $[-80,+140]$\,km\,s$^{-1}$. For a distance of 3.45\,Mpc, we found a total luminosity of 7.04 and 7.00\,K\,km\,s$^{-1}$\,pc$^2$, for NRO and IRAM, respectively. The difference amounts to only 0.6\%. We then compared the flux between the IRAM 30m data cube, the NOEMA+30m combined cube, and the NOEMA-only cube. We here used the same channel spacing of 5\,km\,s$^{-1}$ and the same velocity range as above, but we restricted the field of view to $500''\times500''$ to avoid edge effects. We first regrid the 30m data onto the NOEMA spatial grid, and we smoothed the NOEMA+30m and NOEMA-only cubes to the spatial resolution of the 30m data, that is, $22.5''$. We found a luminosity of 6.20, 6.14, and 0.95\,K\,km\,s$^{-1}$\,pc$^2$ for the IRAM 30m, NOEMA+30m, and NOEMA-only cubes. The flux scales for the combined and single-dish data cubes thus agree within 1\%. Only 15\% of the total flux is recovered with the interferometer.}

From the deconvolved data cubes, we produced a set of derived products including cubes with round beams at a series of fixed angular ($5, 15, 21$\,arcsec) and physical ($90, 150, 500, 1000, 1500$\,pc) resolutions, noise estimates, masks, and accompanying moment maps and uncertainties. The moment maps were produced using a dilated mask approach that follows the strategy presented in \citet{LEROY21c} which was applied to the products delivered for PHANGS--ALMA \citep{LEROY21a}. 
\mod{Line widths were measured using the effective width approach \citep[following][]{HEYER01}, which is often more robust than second-order moment maps: $\sigma = I_{\rm CO}/(\sqrt{2 \pi} T_{\rm peak})$.}
Details are provided in Appendix~\ref{sec:derived-products}. Data cubes and associated data products are publicly available\footnote{\url{https://www.canfar.net/storage/vault/list/phangs/RELEASES/Querejeta_etal_2023}}.

\mod{The conversion from CO(1-0) integrated intensity to molecular gas surface density follows}

\begin{equation}
\Sigma_{\rm mol} = \alpha_{\rm CO}\,I_{\rm CO}\,\cos{i}~,
\label{eq:SigMol}
\end{equation}

\noindent \mod{where $\alpha_{\rm CO}$ represents the conversion factor from CO(1-0) to H$_2$ surface density (including a factor 1.36 to account for helium and heavier elements). The factor $\cos{i}$ corrects the surface densities for inclination.}
Throughout this paper, we apply a fixed $\alpha_{\rm CO} =
4.35\,M_\odot$\,pc$^{-2}$ (K\,km\,s$^{-1}$)$^{-1}$ to IC\,342 whenever
converting from CO intensity to molecular gas mass or surface density. This
value is appropriate for the Milky Way and typical of the disks of massive
spiral galaxies \citep{BOLATTO13B}. We do expect $\alpha_{\rm CO}$ to vary \mod{across IC\,342},
approaching a lower value in the center
\citep[e.g.,][]{DOWNES98,SANDSTROM13,CHIANG21} and to anticorrelate with
the known metallicity gradient of IC\,342 \citep[][and see
Table~\ref{tab:ic342}]{PILYUGIN14}. \mod{Recent dust-based analysis points to an $\alpha_{\rm CO}$ value in the central $\sim$kpc which is ${\sim}4$ times below the Galactic value \citep[I.-D.~Chiang, priv.~comm.;][]{CHIANG23}.
According to \citet{MEIER01} and \citet{PAN14}, in the central few hundred parsec of IC\,342, $\alpha_{\rm CO}$ is $\sim$2-3 times lower than the Galactic value; \citet{PAN14} suggest that the disk of IC\,342 reaches $\alpha_{\rm CO}$ values two times above the Galactic $\alpha_{\rm CO}$ (albeit with significant fluctuations). This means that we can expect up to a factor of ${\sim}5$ bulk difference in $\alpha_{\rm CO}$ between disk and center. Based on this,} we note the cases below where deviation from the adopted constant conversion factor likely influences our results. 

\subsection{Supporting data and measurements}
\label{subsec:supporting}

We used VLA \hi{} 21\,cm line observations from \citet{CHIANG21} to trace neutral atomic gas.
\mod{Specifically, we applied the following equation to transform \hi{} observations to atomic gas surface density, which assumes optically thin 21~cm emission:}

\begin{equation}
    \frac{\Sigma_{\rm atom}}{\rm M_\odot \, pc^{-2}} = 2.0\times10^{-2} \left(\frac{I_{\rm HI}}{\rm K\,km\,s^{-1}}\right) \, \cos{i}~,
    \label{eq:SigHI}
\end{equation}

\noindent \mod{where $\Sigma_{\rm atom}$ includes the 1.36 factor to account for helium and heavier elements. The $\cos{i}$ term again corrects for galaxy inclination.}

We used a \textit{WISE} $3.4\,\mu$m image to trace stellar mass and a \textit{WISE} $22\,\mu$m image to estimate the local recent star formation rate (SFR) surface density \citep{WRIGHT10}. The \textit{WISE} images were processed into an estimate of stellar mass following the procedures in \citet{LEROY19,LEROY21b}. The \textit{WISE} $3.4\,\mu$m image has significant contamination by foreground stars because IC\,342 lies behind the Galactic plane. But since the only application of these data in this paper is large regional averages and radial profiles, we do not expect this to represent a large concern. The $22\,\mu$m image is not heavily affected by the Galactic foreground, and we proceed with the \textit{WISE} $22\,\mu$m alone as our SFR tracer. \mod{We used the following prescription to convert the \textit{WISE} $22\,\mu$m map to SFR:}

\begin{equation}
\frac{\Sigma_{\rm SFR}}{\rm M_\odot \, yr^{-1} \, kpc^{-2}} = 3.8\times10^{-3}\,\left(\frac{I_{22\,{\rm \mu m}}}{\rm MJy\,sr^{-1}}\right) \cos{i}~.
\label{eq:SFR}
\end{equation}

\mod{This prescription follows \citet{LEROY21a} and assumes a Chabrier initial mass function \citep[][]{CHABRIER03}, which agrees with extinction-corrected H$\alpha$ measurements of SFR from the PHANGS--MUSE survey \citep{BELFIORE22} within ${\sim}20{-}30\%$. In this case,} we also use a Mayall MOSAIC H$\alpha$ image kindly provided by K.~Herrmann (private communication) for a visual comparison.

We also made heavy use of an environmental ``mask'' based on the near-IR morphology of the galaxy following \citet[][]{QUEREJETA21}. Appendix~\ref{app:environment} describes the construction of this mask in more detail and explains how we identify distinct ``center,'' \mod{``bar,'' } ``arm,'' and ``inter-arm'' regions.

We aggregated information from the cloud catalogs, SFR estimates, and molecular gas mass estimates into a set of regional property estimates. {This aggregation follows the methods described in \citet{SUN22} and resembles those deployed in \citet{LEROY17A} and \citet{UTOMO18}. Briefly, we calculated the mass-weighted average of cloud properties as well as the aperture average of SFR and molecular gas surface densities in a series of 1.5~kpc hexagonal apertures, which together tile the entire footprint of the NOEMA+IRAM observations. Details about this data aggregation scheme can be found in \citet{SUN22}.} We used these derived products to compare population-averaged estimates of the cloud gravitational free-fall time to the regional molecular gas depletion time. These measurements can be found in Table~\ref{table:hexagons}.

\section{Results}

Figure~\ref{fig:ic342:false-color} shows the molecular gas distribution as mapped by NOEMA+IRAM~\mbox{30-m}, together with the distributions of H$\alpha$ line emission tracing gas ionized by young, massive stars (K.~Herrmann private communication), and neutral atomic gas traced by the 21\,cm line \citep{CHIANG21}. Together, these three tracers reveal an intricate network of multiple spiral arms that emanate from the bright center. The inner arms visible in CO emission and dominated by molecular gas appear prominent inside a galactocentric radius of 5\,kpc. These molecular gas arms connect seamlessly to the outer arms traced by 21\,cm emission and dominated by atomic gas, which can be traced all the way out to radii of ${\sim}20$\,kpc. The ionized gas emission follows these spiral arms, showing a patchy distribution that reflects the presence of many individual 
\ion{H}{II} regions and indicating that massive stars are actively forming along the gas arms. 

Figure~\ref{fig:bigpicture} shows our \chem{CO}{10} survey in more detail. The top left panel illustrates the overall structure of CO emission in the galaxy for the observed field of view. Though the galaxy has been mapped before by lower-resolution, lower-sensitivity, or smaller field-of-view observations \citep[including][]{CROSTHWAITE01,HELFER03,KUNO07,HIROTA11}, the NOEMA map offers the sharpest, most complete view to date of the CO distribution in IC\,342. The peak intensity map in Fig.~\ref{fig:bigpicture} shows multiple spiral arms, fine inter-arm filaments, and a bright, elongated inner structure, which is reminiscent of gas lanes along a stellar bar and encompasses the nuclear starburst \citep[e.g.,][]{DOWNES92,SCHINNERER03,MEIER05}. The spiral arm east of the center has a significantly higher contrast similar to the (eastern) \hi{} spiral arms further out in the disk (see Fig.~\ref{fig:ic342:false-color}). We speculate that this might be related to a mild interaction with a neighboring galaxy, which has also been evoked to explain the warping of the outer \hi{} disk \citep[e.g.,][]{ROTS79,CROSTHWAITE00,STIL05}. The CO velocity field shown in the top right panel \mod{follows a gentle S-shape which is} consistent with \mod{streaming motions associated with a bar potential on top of a} regularly rotating gas disk.

The bottom row of Fig.~\ref{fig:bigpicture} places our new CO survey in context. We combine the distribution of the atomic gas from \citet{CHIANG21} with our new molecular gas map at a matched resolution of ${\sim}350$\,pc to show the total neutral gas surface density distribution (\textit{left panel}) and to construct radial profiles of atomic gas, molecular gas, and stellar surface density (\textit{right panel}). The effect of changing angular resolution is particularly apparent for the central bright molecular gas distribution where the position angle changes from a small offset to the east at ${\sim}4\arcsec$ resolution to a small western offset at $20\arcsec$ resolution. This change is caused by lower surface brightness material outside the central $60\arcsec$ which becomes apparent at the lower angular resolution. This might be \mod{a consequence of the bar in IC\,342} (see Sect.\,\ref{subsec:bar}).

The radial profiles show the basic distribution of the cold ISM in IC\,342. Within 6\,kpc, the molecular gas and stellar mass \mod{profiles have a nearly constant offset in logarithmic scale,} implying a molecular gas to stellar fraction $\Sigma_\emr{mol}/(\Sigma_\star+\Sigma_\emr{mol}) \approx 0.20{-}0.25$ \mod{for a constant Galactic $\alpha_{\rm CO}$; if we assume a 2-3 times lower $\alpha_{\rm CO}$ in the center and twice higher $\alpha_{\rm CO}$ in the disk, as suggested by \citet{PAN14}, the molecular to stellar fraction would drop to ${\sim}10\%$ for the center and reach ${\sim}40\%$ throughout the disk.} The molecular gas surface density exceeds the atomic gas surface density out to at least $r_{\rm gal} = 6.7$\,kpc.
Outside approximately this radius, the atomic gas dominates the cold ISM. As shown in Fig.~\ref{fig:ic342:false-color} and~\ref{fig:bigpicture}, the structure of the molecular gas appears to connect smoothly with the atomic gas, consistent with a density- or pressure-dependent phase change. This large-scale picture agrees well with previous low-resolution work by \citet{CROSTHWAITE01} \mod{and \citet{PAN14}}. However, in contrast to previous work, our map captures the sharp concentration of gas at the galaxy center ($r_{\rm gal} < 1$\,kpc) and resolves the CO emission from the galaxy into discrete molecular clouds. We explore the implications of these higher resolution observations in the next sections.

\subsection{Giant molecular clouds in \texorpdfstring{IC\,342}{IC342}}
\label{GMCs}

\FigGMCs{} %

We used the python implementation of the \texttt{CPROPS} algorithm \citep{ROSOLOWSKY06} to generate a catalog of molecular clouds in IC\,342. We apply the algorithm to the $90\,\mathrm{pc}$ 
and $5$\,km\,s$^{-1}$ resolution cube and use the same methods developed by \citet{ROSOLOWSKY21} for the PHANGS--ALMA survey; see Appendix~\ref{app:clouds} for details. These methods yield radius~($R$), line width~($\sigma$), and mass~($M$) for each cloud. 
\mod{We relied on the extrapolated and deconvolved GMC properties that account for the spatial and spectral response of NOEMA. When computing cloud radii, we followed $R = \eta \sqrt{\sigma_{\rm maj,d} \sigma_{\rm min,d}}$ with $\eta = 1.18$ as in \citet{ROSOLOWSKY21}.} We adopted a fixed, Galactic $\alpha_{\rm CO}$ as stated above.

The \texttt{CPROPS} catalog contains \ngmc{} clouds with masses between $10^5$ and $4\times 10^7\,M_\odot$ \mod{(out of which 602 have radius measurements)}. \mod{The catalog is publicly available\footnote{\url{https://www.canfar.net/storage/vault/list/phangs/RELEASES/Querejeta_etal_2023}} and an excerpt showing the first ten rows can be seen in Table~\ref{table:GMCs}.} Given the $\theta = 90$\,pc resolution and this mass range, these objects likely span the range between giant molecular clouds and giant molecular associations. As we see below, they appear approximately gravitationally bound. Fig.~\ref{fig:clouds} visualizes their sizes and locations in the galaxy, with the clouds colored according to their dynamical environment (see Appendix~\ref{app:environment}). 

The top right and bottom left panels of Fig.~\ref{fig:clouds} illustrate the dynamical state of these molecular clouds using plots of $\sigma^2/R$ versus $\Sigma_{\rm mol}\equiv M/(2\pi R^2)$ space. Following \citet{HEYER09} and \citet{FIELD11}, variants of this plot have become a standard way to assess the properties of a molecular cloud population \citep[e.g., see][]{LEROY15A,SUN18,SUN20B,ROSOLOWSKY21}. 
\mod{In the log--log plot of $\sigma$ as a function of $\Sigma_{\rm mol}$ (bottom right panel), clouds with fixed virial parameter $\alpha_{\rm vir} = 2 {\rm KE} / {\rm UE}$, that is, a fixed ratio of kinetic energy (KE) to self-gravitating potential energy (UE), follow a straight line with slope of~$1$.}
Conversely, clouds with fixed turbulent pressure follow straight lines with slope of~$-1$ \citep{SUN18,SUN20B}. Figure~\ref{fig:clouds} shows the IC\,342 clouds, separated by environment (as illustrated in the top left panel). As a reference we show a large set of GMC properties derived at identical spatial resolution and similar sensitivity from PHANGS--ALMA \citep[A.~Hughes et al., in preparation]{ROSOLOWSKY21}. {We note that PHANGS--ALMA relies on the \mbox{CO(2-1)} transition, while the IC\,342 observations presented here mapped \mbox{CO(1-0)}; the ratio of both lines, $R_{21}$, shows some mild variations within galaxies \citep[e.g.,][]{DENBROK21,LEROY21c}.} The upper right panel shows individual clouds, while the lower left panel shows running medians in $\sigma^2/R$ as a function of $\Sigma_{\rm mol}$. In the lower left panel, we plot the mass-weighted median and $16{-}84$\% range in surface density for each dynamical environment in IC\,342.

In the bottom right panel of Fig.~\ref{fig:clouds}, we show a complementary view of the molecular ISM that captures similar information to the top right and bottom left panels. Here, we use a line-of-sight based approach following \citet{SUN18, SUN20B}. We again take the $\theta=90$\,pc data and now measure the intensity and line width along each line of sight. We sample the map using a hexagonal grid with spacing equal to the beam FWHM and consider only lines of sight where CO is detected over at least three $5$\,km\,s$^{-1}$ channels. Then we show $\sigma$ at this fixed spatial scale, which will be analogous to $\sigma^2/R$ for the clouds, as a function of $\Sigma_{\rm mol}$, also at the fixed scale. Again we show results for PHANGS--ALMA, now from \citet{SUN20B} for reference. Here we also plot the mass-weighted median and $16{-}84$\% range for each different environment. \mod{We plot typical error bars as the median fractional error associated with the NOEMA intensity map (14\%) and linewidth map (18\%) for $\Sigma_{\rm mol}$ and $\sigma$, respectively; we use a bootstrap approach (with $N=1000$ iterations) for the uncertainty in $\sigma^2/R$, taking the difference between the CPROPS radius with and without deconvolution as a representative error bar for $R$.}

The two approaches and all three panels tell a consistent story. The clouds seen in IC\,342 have properties broadly consistent with other clouds seen throughout the local galaxy population. Some of the flattening in line width at low surface density reflects the $2\times$ coarser velocity resolution of the IC\,342 data compared to PHANGS--ALMA (in gray). If we consider the dynamical state in only the high surface density, $\Sigma_{\rm mol} > 100\,M_\odot$\,pc$^{-2}$ clouds, where the CO line is typically well-resolved by NOEMA, then we find a median virial parameter 
\mod{$\alpha_{\rm vir} = 2 {\rm KE} / {\rm UE} \approx 1.2$ and $1.5$ in the arm and inter-arm regions, respectively. The median virial parameter is higher ($\alpha_{\rm vir} \approx 2.0$) within the bar region, but then drops again for clouds in the very center ($\alpha_{\rm vir} \approx 1.4$). In any case, we find a range of virial parameters across environments (the global 16th-84th percentile range goes from $0.8$ to $4.0$, with a median $\alpha_{\rm vir} \approx 1.6$).}
As the figure shows, these arm and inter-arm results are well in-line with the PHANGS--ALMA clouds. \mod{The bar, on the other hand, shows enhanced $\sigma^2/R$ compared to arm or inter-arm values at matched molecular gas surface density.} For the center, we note two subtleties. First, the highest surface density regions associated with the most massive clouds or pixels in the lower panels of Fig.~\ref{fig:clouds} actually have $\alpha_{\rm vir}$ lower than~$2.0$, indicated by lower $\sigma$ at fixed $\Sigma_{\rm mol}$. In particular, the galaxy center itself appears, at our 90\,pc resolution, roughly virialized, though we caution that 90\,pc is quite coarse for such a study \citep[e.g., see][]{SCHINNERER06}. Second, we caution that, as mentioned above, the center of IC\,342 likely has a lower $\alpha_{\rm CO}$ than the Galactic value that we adopt \citep{ISRAEL20,CHIANG21}. This will have the effect of moving the center points in all panels to the left, increasing their virial parameter by decreasing the surface density and thus the gravitational potential.

The surface density and line width increase moving from the inter-arm to the arm \mod{or bar, and} then to the center regions. Weighting by mass, the \mod{median $\Sigma_{\rm mol}$ is ${\sim}80, 140, 160$, and $1100\,M_\odot$\,pc$^{-2}$ for inter-arm, arm, bar, and center clouds, respectively, while the corresponding median line widths are $6.6$, $7.6$, $9.7$ and $18.4$\,km\,s$^{-1}$.} 
The beam-by-beam statistics show similar trends, with mass-weighted median $\Sigma_{\rm mol}$ of 
\mod{$50$, $80$, $90$, and $820\,M_\odot$\,pc$^{-1}$ and mass-weighted median line widths of $5.2$, $6.2$, $7.7$, and $17.2$\,km\,s$^{-1}$ in the inter-arm, arm, bar, and center regions. According to \citet{PAN14}, $\alpha_{\rm CO}$ could be 2-3 times lower in what we define as the center region (innermost $\sim$kpc), which would lead to more moderate center surface densities (${\sim}500$ and ${\sim}300\,M_\odot$\,pc$^{-2}$ for GMCs and beam-by-beam, respectively).}
\mod{A two-sided Kolmogorov-Smirnov test confirms that the arm-interarm differences are statistically significant ($p$-value $<0.1$\%). At a similar level, the center is clearly different from the rest. The bar surface densities are not statistically distinct from the spiral arm values ($p$-value 36.8\% for GMCs; 2.2\% pixel-by-pixel), but bars do deviate from spiral arms in terms of $\sigma^2/R$ and $\sigma$ ($p$-value $<0.1$\%).}

Overall, these trends show an increase in surface density and line width moving from inter-arm to arm regions but with an approximately similar dynamical state. The center shows an increased surface density and line width. Many clouds in the central region, though not in the immediate center itself, show high line width relative to their surface density and evidence of broadening by forces other than self-gravity. A depressed $\alpha_\emr{CO}$ value would even emphasize this effect for the center. These results agree well with those found for larger populations of more distant galaxies in PHANGS--ALMA \citep{SUN20B,ROSOLOWSKY21}.

\subsection{Star formation efficiency per free-fall time}

\FigSFE{} %

The rate at which stars form out of gas is often claimed to be primarily driven by gas density, which determines the local free-fall time, $\tau_\mathrm{ff}$ \citep[e.g., see reviews in][]{LADA03,KENNICUTT12,FEDERRATH13,PADOAN14}. Therefore, models of star formation often treat $\tau_\mathrm{ff}$ as a characteristic timescale for star formation, and express the fraction of gas converted to stars per gravitational free-fall time via $\epsilon_\mathrm{ff} = \mathrm{SFR} / (M_\mathrm{mol} / \tau_\mathrm{ff})$, a.k.a. the ``efficiency per free-fall time'' \citep[e.g.,][]{MCKEE07,KRUMHOLZ19}. Some work has suggested $\epsilon_{\rm ff}$ to be approximately constant across a wide range of scales and systems, so that density variations represent a main physical driver of variations in the molecular gas depletion time \citep[e.g., see][]{KRUMHOLZ07B,KRUMHOLZ19}. In other work, $\epsilon_{\rm ff}$ represents a key quantity but varies depending on the local physical state of the gas \citep[e.g.,][]{UTOMO18,SUN23}.

We estimated the molecular cloud free-fall time $\tau_\mathrm{ff}$ in IC\,342 from our high-resolution CO data. Our \texttt{CPROPS} analysis provides measurements of mass and size for each identified molecular cloud. From these measurements, we estimated the mean volume density via $\rho = 3M/4\pi R^3$ and the associated gravitational free-fall time via $\tau_\mathrm{ff} = \sqrt{3\pi/(32G\rho)}$. We then averaged over all clouds in each $1.5$\,kpc-diameter hexagonal region of IC\,342, calculating the mass-weighted mean reciprocal of $\tau_\mathrm{ff}$ via
\begin{equation}
\langle\tau_\mathrm{ff}\rangle^{-1}
= \frac{\sum_i M_i \tau_\mathrm{ff,i}^{-1}}{\sum_i M_i}~,
\label{eq:tau_ff}
\end{equation}
\noindent where $i$ is the index of clouds in that kpc-sized aperture. 
\mod{We follow \citet{UTOMO18} and \citet{SUN22,SUN23} in calculating the mass-weighted harmonic mean of $\tau_\mathrm{ff}$, which ensures that the result is representative of the free-fall time for the entire cloud population in each hexagonal aperture.}
This ensures that $\langle\tau_\mathrm{ff}\rangle^{-1}$ is the correct quantity to compare to $\tau_{\rm dep}^{\rm mol}$ to test the hypothesis of fixed $\epsilon_{\rm ff}$. \mod{We estimated typical error bars on our measurements following a bootstrap approach with $N=1000$ iterations. Within each hexagonal aperture, we perturbed the surface density and radius of each cloud following a Gaussian with standard deviation equal to 14\% of each cloud's surface density and 16\% of its radius, which are the typical error bars that we have derived across our sample. In each iteration, we recomputed $\rho$ and $\tau_{\rm ff}$ for each cloud with these perturbed values, and calculated the mass-weighted average of $\tau_{\rm ff}^{-1}$ within each hexagonal aperture. Based on these $1000$ iterations, we obtained the fractional error on $\tau_{\rm ff}$ for each aperture, and computed the median across all apertures, which is 6\%. For $\tau_\mathrm{dep}^\mathrm{mol}$, we assume the same 14\% typical uncertainty on $\Sigma_{\rm mol}$ and take the median difference between the SFR calculated from WISE~4 and WISE~3 (17\%) as a representative error bar. These combine to yield a typical error bar of 22\% for $\tau_\mathrm{dep}^\mathrm{mol}$.}

The median $\langle\tau_\mathrm{ff}\rangle$ in our sample of regions is $11$\,Myr with a $16{-}84$\% range of $8.5{-}14.8$\,Myr.
\mod{This is slightly longer than the median value of $\tau_\mathrm{ff} = 6.4 ^{+3.5}_{-2.5}$ found for other galaxies at this scale by \citet{ROSOLOWSKY21}, but still in agreement within $2\sigma$ given the measured galaxy-to-galaxy variation.}
For comparison, \citet{KIM21}, also using these data, made a statistical estimate of the cloud lifetime in IC\,342 and found ${\sim}20$\,Myr. This implies that clouds live ${\sim}2\tau_{\rm ff}$ in this galaxy, in good agreement with the results of \citet{KRUIJSSEN19}, \citet{CHEVANCE20}, and \citet{KIM22} for a broader sample of galaxies.

In addition to $\langle\tau_\mathrm{ff}\rangle$, we also calculated the integrated molecular gas depletion time, $\tau_\mathrm{dep}^\mathrm{mol}$,  in each region via 
\begin{equation}
\tau_\mathrm{dep}^\mathrm{mol} = \frac{M_\mathrm{mol}}{\mathrm{SFR}}~.
\label{eq:tau_dep}
\end{equation}
Here, $M_\mathrm{mol}$ and $\mathrm{SFR}$ are the integrated molecular gas mass and star formation rate in that aperture
\mod{(measured from \textit{WISE} $22\,\mu$m using Eq.\,\ref{eq:SFR}).}
A high $\tau_\mathrm{dep}^\mathrm{mol}$ corresponds to a low rate of star formation per unit molecular gas and implies that star formation will take a long time to consume the available gas reservoir.

We combined $\langle\tau_\mathrm{ff}\rangle$ and $\tau_\mathrm{dep}^\mathrm{mol}$ to estimate the mean efficiency per free-fall time in that aperture via $\epsilon_{\rm ff} = \langle\tau_\mathrm{ff}\rangle / \tau_\mathrm{dep}^\mathrm{mol}$. By using $1.5$\,kpc-sized apertures, we average over many clouds at different evolutionary stages, which makes our results robust to stochasticity and the evolutionary state of individual clouds. This approach follows similar work by \citet{LEROY17A}, \citet{UTOMO18} and \citet{SCHRUBA19}, and the hexagonal sampling scheme and measurements follow \citet{SUN20}.

In Fig.~\ref{fig:eff}, we compare $\langle\tau_\mathrm{ff}\rangle$ and $\tau_{\rm dep}^{\rm mol}$ for each $1.5$\,kpc region. Each hexagon symbol shows results for one aperture, with the points color-coded according to the deprojected galactocentric radius at the aperture center. The red line shows the expectation for $\tau_{\rm dep} = \langle\tau_\mathrm{ff}\rangle / \epsilon_{\rm ff}$ at the median measured $\epsilon_{\rm ff}$ and the pink region shows the $16{-}84$\% range of estimated $\epsilon_{\rm ff}$. If $\epsilon_{\rm ff}$ is indeed constant and if density variations drive variations in $\tau_{\rm dep}^{\rm mol}$ then we would expect the data to follow that red line.

For our $90$\,pc resolution cloud catalog and using \chem{CO}{10} with a fixed conversion factor and our \textit{WISE}~22$\,\mu$m-based SFR estimates, we find a median of $\epsilon_{\rm ff} = 0.45\%$ with a $16{-}84$\% range of $0.33{-}0.71\%$. This value is near the frequently invoked value of $0.5\%$ \citep[e.g.,][]{MCKEE07} and broadly consistent with  previous estimates in the literature \citep[e.g.,][]{VUTISALCHAVUAKUL16,BARNES17,LEROY17A,UTOMO18,SCHRUBA19}. In detail, the value is towards the low end of the distribution for PHANGS--ALMA galaxies found by \citet{UTOMO18} and intermediate between their $0.7\%$ mean value and the $0.3\%$ found at $40$\,pc resolution for PAWS by \citet{LEROY17A}. All of these values tend to be somewhat lower than those found for Galactic objects.

We can wonder if our IC\,342 measurements support the idea of a fixed $\epsilon_{\rm ff}$. On one hand, within the disk of the galaxy, that is, excluding the central aperture, there does not appear to be any significant relation between $\langle\tau_\mathrm{ff}\rangle$ and $\tau_{\rm dep}^{\rm mol}$. The Spearman's rank correlation relating the two is about $-0.14$, which goes opposite the predicted direction for a constant $\epsilon_{\rm ff}$, and the relationship has an insignificant $p$-value of $0.26$. The disk of IC\,342 appears to exhibit a similar value of $\epsilon_{\rm ff}$ to other cases, but any internal changes in $\tau_{\rm dep}^{\rm mol}$ do not show clear evidence of being driven by density variations.

On the other hand, the contrast between the central $1.5$\,kpc averaged measurement, indicated by a bright yellow hexagon, and the cloud of regions in the disk does offer some support for the ideas that $\epsilon_{\rm ff}$ is fixed and that density variations drive $\tau_{\rm dep}^{\rm mol}$ variations. Unfortunately, for the center, $\alpha_{\rm CO}$ complicates the picture. We have adopted a single fixed $\alpha_{\rm CO}$ for Fig.~\ref{fig:eff} but expect that the center, at least, shows a depressed $\alpha_{\rm CO}$. For instance, \citet{MEIER11}, \citet{PAN14}, \citet{ISRAEL20}, and \citet{CHIANG21}, all find low $\alpha_{\rm CO}$ for the inner part of IC\,342. This can have a dramatic effect on $\epsilon_{\rm ff}$, which we illustrate by also plotting results for the central $1.5$\,kpc of IC\,342 using a ``starburst'' $\alpha_{\rm CO} = 0.8\,M_\odot$\,pc$^{-2}$ (K\,km\,s$^{-1}$)$^{-1}$ conversion factor \citep{DOWNES98,BOLATTO13B}.
\mod{We do not expect $\alpha_{\rm CO}$ to be necessarily this low in the center of IC\,342, 
but this implies a difference of a factor of ${\sim}5$ between center and disk, which is probably realistic. For example, \citet{PAN14} found a central $\alpha_{\rm CO}$ 2-3 below the Galactic value, while for the disk $\alpha_{\rm CO}$ was found to be typically two times above the Galactic value, amounting to a similar difference of a factor ${\sim}5$.
Applying this five times lower $\alpha_{\rm CO}$ in the center} implies a significantly {higher} $\epsilon_{\rm ff}$ for the central region, ${\sim}4.2\%$ instead of $0.3\%$. This suggests that $\epsilon_{\rm ff}$ varies between the disk and the central region \mod{and} highlights the importance of accurate $\alpha_{\rm CO}$ to these estimates.


\section{Discussion: Central molecular gas reservoir and nuclear star cluster}
\label{subsec:bar}

\FigZoom{}

Figure~\ref{fig:zoom} shows three zooms of the IC\,342 center after deprojection of its inclination on the plane of sky. Several properties of IC\,342's central region are similar to those in our Milky Way: the high molecular gas concentration, its geometry, and the massive nuclear star cluster \citep[NSC, e.g.,][]{SCHINNERER03,NEUMAYER20}. A striking feature of the new CO map is the sharp increase in molecular gas surface density in the central ${\sim}1$\,kpc (see Fig.~\ref{fig:bigpicture}). The average molecular mass surface density in the central 1.5\,kpc measured at 90\,pc resolution is about $600\,M_{\odot}\,{\rm pc}^{-2}$ (for a Galactic $\alpha_{\rm CO}$), that is, a factor of $10\times$ higher than in the immediate surroundings. Such sharp contrasts in CO surface brightness between center and disk are only seen in nearby galaxies that host a prominent stellar bar \citep[see, e.g., CO atlas from PHANGS--ALMA,][]{LEROY21b}.

\mod{The non-axisymmetric gas morphology in the center appear related to the underlying dynamical structure in IC\,342. However, the existence of a stellar bar in IC\,342 has been debated in the literature.}
\citet{BUTA99} did not discuss a potential bar, but \mod{they carried out ellipse fitting of $V$ and $I$ band images that shows changes in position angle and ellipticity which we find suggestive of a large-scale bar component.}
Based on ellipse fitting of deep near infrared images, \citet{FATHI09} reported a large-scale stellar bar of $307\arcsec$ (or ${\sim}5.1$\,kpc) semi-major axis oriented roughly north--south, while \citet{HERNANDEZ05} found the bar to be not well-defined. \citet{CROSTHWAITE00} reported evidence for a ``fat'' or ``boxy'' bar potential based on the analysis of their \hi{} kinematic data.

\mod{We examined high-quality NIR imaging from \textit{Spitzer} IRAC 3.6\,$\mu$m and found clear evidence for the presence of a stellar bar}
with semi-major length of $120\arcsec$ (or ${\sim}2$\,kpc) and
\mod{an orientation of $\mathrm{P.A.}=160^\circ$ 
(Appendix~\ref{app:environment})}. 
\mod{In addition to this photometric evidence, the presence of a bar is also suggested by CO morphology}
in the central $60\arcsec$ \citep[see for a summary and sketch][]{MEIER05}. The distribution of the CO line 
emission shows two \mod{slightly curved} gas lanes ending towards the center in a ring-like shape which coincides with a ${\sim}100$\,pc diameter star-forming \mod{pseudoring (or nuclear spiral)}, and shock tracers appear preferentially on the leading sides of the lanes close to where they connect to the ring \mod{(a morphology that is clearly revealed by our data, see Fig.\,\ref{fig:zoom}.)}. 
\mod{In the whole PHANGS--ALMA CO morphology catalog, such a setup is only observed in barred galaxies \citep{STUBER23}.}
\mod{Moreover, CO morphology suggests that it is a relatively weak bar, because the bar lanes are curved and do not extend all the way out to the bar ends \citep[e.g.,][]{ATHANASSOULA92,COMERON09}. The CO kinematics are also consistent with the gas responding to an underlying non-axisymmetric (bar) potential.}
\mod{The presence of bar lanes} is also supported by observations of the magnetic field \citep{BECK15}. 
Finally, the NSC in IC\,342 is at the same time extremely massive and compact -- consistent with a picture where \mod{bar-driven gas inflow triggers} in situ star formation \mod{that drives the stellar densities in NSCs to extremely high values \citep{NEUMAYER20,PECHETTI20}.}

\mod{In this paper, we have presented measurements showing elevated gas velocity dispersion within the center environment (Sect.\,\ref{GMCs}). The current central gas concentration and bar lanes, which are connected to a pseudoring (or nuclear spiral), can be interpreted as a consequence of that high velocity dispersion. Indeed, theoretical studies suggest that local gas properties such as the effective sound speed play a key role in determining the gas morphology within bars, beyond the existence or not of an inner Lindblad resonance \citep[ILR; e.g.,][]{ENGLMAIER97,ENGLMAIER00,SORMANI18}. The expectation is that warmer gas with high velocity dispersion (higher sound speed) will result in a more open nuclear spiral, which also critically depends on the central mass concentration.}

\mod{Simulations show that the morphology of these central gas structures varies quickly over time \citep[e.g.,][]{EMSELLEM15,RENAUD15,SORMANI20}. Therefore, the present configuration is probably not long-lived and was different in the past. We speculate that there might have been a stable nuclear ring associated with the bar in the past, which prevented gas inflow; the gas in this ring could have experienced dynamical and thermal heating due to feedback, giving way to the current open spiral morphology. This could signify that gas has only recently been able to reach the center, where it could contribute to the growth of the massive NSC. This process could have occurred intermittently, tied to rapid changes in the central gas morphology (i.e., from ring to open spiral and back again) that lead to periodic gas inflow to the center. In this case, bar-driven inflow episodes like this over the course of IC342's recent history would be responsible for building the NSC. Indeed, the nuclei of barred galaxies typically host much more elevated gas surface densities than unbarred galaxies \citep[e.g.,][]{SAKAMOTO99,SUN20}. It should be kept in mind, however, that it is difficult to determine precisely the past evolutionary history from the present gas distribution \citep[][]{RENAUD15,SORMANI20}.}


\section{Conclusions}

We have presented a new, large, sensitive, ${\sim}1000$-pointing \chem{CO}{10} line mosaic of the very nearby galaxy IC\,342, carried out using the IRAM NOEMA and \mbox{30-m} telescopes. The survey covers the inner ${\sim}11 \times 11$\,kpc$^2$ region, which corresponds to roughly $1.5$ effective radii and extends out to near the radius where the galaxy becomes atomic gas dominated. With a linear resolution of ${\sim}60$\,pc, our beam is about the size of a massive giant molecular cloud. We resolve the galaxy into a series of bright spiral arms, inter-arm filaments, and the well-known bright central molecular structure which is 
\mod{consistent with} gas flowing along a stellar bar. The well-known bright starburst center of IC\,342 appears prominent in all of our data products.

We applied the \texttt{CPROPS} algorithm to identify ${>}600$ massive molecular objects that are either molecular clouds or molecular associations. We did this at $90$\,pc resolution, so that IC\,342 can be placed in the context of a large set of molecular cloud properties recently measured by \mod{PHANGS--}ALMA. We also conducted a parallel ``beam-wise'' analysis in which we estimated the line width and molecular surface density for each independent line of sight where CO is detected over at least three consecutive channels. Both analyses highlight that the surface density and line width increase from the inter-arm region to the arm \mod{and bar} region, and gas in the center of the galaxy has the highest surface density and line width. 
\mod{For clouds, we find a mass-weighted median $\Sigma_{\rm mol} \approx 80, 140, 160$, and $1100\,M_\odot$\,pc$^{-2}$, and $\sigma_{\rm CO} \approx 6.6$, $7.6$, $9.7$ and $18.4$\,km\,s$^{-1}$ for inter-arm, arm, bar, and center, respectively. Our beam-by-beam statistics yield similar results, $\Sigma_{\rm mol} \approx 50, 80, 90$, and $820\,M_\odot$\,pc$^{-1}$, and $\sigma_{\rm CO} \approx 5.2, 6.2, 7.7$, and $17.2$\,km\,s$^{-1}$, respectively. A $2$-$3$ times lower $\alpha_{\rm CO}$ in the center yields median surface densities of ${\sim}500$ and ${\sim}300\,M_\odot$\,pc$^{-2}$ for GMCs and beam-by-beam, respectively.}
Moreover, \mod{for a Galactic $\alpha_{\rm CO}$} clouds appear to be in rough virial equilibrium in both \mod{the cloud-based and ``beam-wise'' analysis}. The clouds in the \mod{bar and} central region show elevated line widths, consistent with dynamical broadening by forces other than self-gravity and in good agreement with previous observations of galaxy centers using both techniques. If we had adopted a lower conversion factor, as is likely appropriate for the inner region, the dynamical state of the gas in the center would depart even more from simple virial equilibrium where turbulence balances self-gravity.

We used the cloud properties measured at $90$\,pc resolution to estimate the density, gravitational free-fall times, and star formation efficiency per free-fall time. We find median $\langle \tau_{\rm ff}\rangle \approx 11$\,Myr and $\epsilon_{\rm ff} = 0.45\%$ on average with a $16{-}84$\% range of $0.33{-}0.71\%$. This is in agreement with literature estimates of this quantity, though towards the low end of such estimates. The center shows shorter molecular gas depletion time and gravitational free-fall time. Whether it has the same $\epsilon_{\rm ff}$ compared to the disk depends sensitively on the adopted $\alpha_{\rm CO}$. Combining these values with recent statistical estimates of the GMC lifetime in IC\,342 implies that clouds in the galaxy live for typically ${\sim}2$ times the free-fall time. 

\mod{Given the proximity of IC\,342, the CO observations and GMC catalog presented in this paper could have a wide range of applications for studies of the nearby universe.}
The data, along with our data products, are publicly available.


\begin{acknowledgements}

\mod{We thank the anonymous referee for a very constructive report.}

We gratefully acknowledge support from the NOEMA observatory staff.

MQ acknowledges support from the Spanish grant PID2019-106027GA-C44, funded by MCIN/AEI/10.13039/501100011033.

JP acknowledges support by the Programme National ``Physique et Chimie du Milieu Interstellaire'' (PCMI) of CNRS/INSU with INC/INP, co-funded by CEA and CNES.

ES and TGW acknowledge funding from the European Research Council (ERC) under the European Union’s Horizon 2020 research and innovation programme (grant agreement No. 694343).

JS acknowledges support by the Natural Sciences and Engineering Research Council of Canada (NSERC) through a Canadian Institute for Theoretical Astrophysics (CITA) National Fellowship.

MC and JMDK gratefully acknowledge funding from the DFG through an Emmy Noether Grant (grant number KR4801/1-1), as well as from the European Research Council (ERC) under the European Union's Horizon 2020 research and innovation programme via the ERC Starting Grant MUSTANG (grant agreement number 714907). MC gratefully acknowledges funding from the DFG through an Emmy Noether Grant (grant number CH2137/1-1). COOL Research DAO is a Decentralised Autonomous Organisation supporting research in astrophysics aimed at uncovering our cosmic origins. MC, JK, and JMDK gratefully acknowledge funding from the German Research Foundation (DFG) through the DFG Sachbeihilfe (grant number KR4801/2-1). CE gratefully acknowledges funding from the Deutsche Forschungsgemeinschaft (DFG) Sachbeihilfe, grant number BI1546/3-1. KK gratefully acknowledges funding from the German Research Foundation (DFG) in the form of an Emmy Noether Research Group (grant number KR4598/2-1).

RSK acknowledges financial support from the European Research Council (ERC) via the ERC Synergy Grant "ECOGAL" (grant 855130) as well as from the Heidelberg cluster of excellence EXC 2181 (Project-ID 390900948) ”STRUCTURES” funded by the German Excellence Strategy. RSK also thanks DFG for support via the collaborative research center (SFB 881, Project-ID 138713538) ”The Milky Way System” (subprojects A1, B1, B2, and B8). 

\end{acknowledgements}

\bibliographystyle{aa} %
\bibliography{ms.bib}

\begin{appendix}

\section{NOEMA observations and reduction}
\label{sec:obs+red}

\begin{table*}
  \centering
  \caption{NOEMA and IRAM~\mbox{30-m} CO(1-0) observations at 115.271202\,GHz}
  \begin{tabular}{lccccc}
    \hline
    \hline
    Data set                              & Project time      & Field of View         & Beam                             & Resolution\tablefootmark{a} & RMS noise\tablefootmark{a}  \\
                                          & hours             & arcmin                & arcsec                           & \emr{km\,s^{-1}}            & K \\
    \hline
    NOEMA\tablefootmark{b}$+$\mbox{30-m}  & $120+43$          & $10.75 \times 10.75$  & $4\times3.25$ at $91^\circ$      & 5.0 &  0.114 \\
    NOEMA\tablefootmark{b}                & 120               & $10.75 \times 10.75$  & $4\times3.25$ at $91^\circ$      & 5.0 &  0.119 \\
    IRAM~\mbox{30-m}                      & 43                & $10.75 \times 10.75$  & $22.5$                           & 5.0 &  0.058 \\
    \hline
  \end{tabular}
  \tablefoot{%
    \tablefoottext{a}{The spectral resolutions and associated RMS noise are given for the best spectral resolution of our products, close to the WIDEX channel spacing. The channel spacing achived at the IRAM-30m was about 10 times smaller.}
    \tablefoottext{b}{NOEMA had 8 antennas in 2016. With 12 antennas and the On-The-Fly observing mode, the same data quality would have been achieved with 26 hours of NOEMA observatory time.}}
  \label{tab:obs}
\end{table*}

\begin{table*}[t!]
\begin{center}
\caption[h!]{\mod{Example catalog of GMC properties of IC\,342. The full table is available online.}
}
\begin{tabular}{lccccccccc}
\hline
\hline 
Cloud & RA & Dec & $v_{\rm LSR}$ & $R$ & PA & $e$ & $\sigma_{v}$ & $M_{\rm CO}$  & $M_{\rm vir}$ \\
 & (J2000) & (J2000)  & (km\,s$^{-1}$) & (pc) & ($^\circ$) & & (km~s$^{-1}$) & ($10^6$\,M$_\odot$) & ($10^6$\,M$_\odot$) \\
\hline
  1 &   56.57360 &  68.029582 &    -55.7 &      --- &   59 &    --- &    6.8 &    0.4 &    ---  \\
  2 &   56.57038 &  68.033543 &    -56.1 &    106.9 &    8 &   0.88 &    7.5 &    2.6 &   10.2  \\
  3 &   56.49006 &  68.052340 &    -49.2 &     66.8 &  174 &   0.42 &    6.7 &    1.2 &    5.0  \\
  4 &   56.55737 &  68.010278 &    -51.2 &     91.3 &  147 &   0.38 &    5.8 &    2.2 &    5.2  \\
  5 &   56.56855 &  68.011551 &    -51.7 &     62.4 &   22 &   0.82 &    4.8 &    1.2 &    2.4  \\
  6 &   56.56769 &  68.017141 &    -53.3 &    112.6 &   20 &   0.72 &    7.0 &    1.2 &    9.4  \\
  7 &   56.52838 &  68.030475 &    -59.5 &     87.0 &   24 &   0.85 &   11.3 &    1.3 &   18.6  \\
  8 &   56.51868 &  68.031299 &    -62.6 &      --- &   34 &    --- &    6.8 &    0.3 &    ---  \\
  9 &   56.54947 &  68.024618 &    -55.2 &    124.7 &   44 &   0.91 &    3.7 &    1.7 &    2.9  \\
10 &   56.54147 &  68.030874 &    -51.6 &    105.8 &  125 &   0.58 &    8.3 &    3.2 &   12.4  \\
\hline
\end{tabular}
\label{table:GMCs}
\end{center}
\tablefoot{The columns correspond to the cloud coordinates, velocity centroid, deconvolved cloud radius, position angle of the cloud major axis, cloud eccentricity, line width, and cloud luminous and virial mass. The cloud eccentricity is calculated as $e = \sqrt{1 - \sigma_{\rm min,d}^2 / \sigma_{\rm maj,d}^2}$. $M_{\rm CO}$ is obtained as the cloud CO luminosity (flux in K\,km\,s$^{-1}$\,pc$^{2}$) multiplied by our assumed Galactic CO-to-H$_2$ conversion factor. The virial mass is estimated as $M_{\rm vir} = 5 \sigma_v^2 R_{\rm 3D} / G$.}
\end{table*}

\begin{table*}[t!]
\begin{center}
\caption[h!]{\mod{Example table of free-fall time and depletion time in IC\,342 (Fig.\,\ref{fig:eff}). The full table is available online.}
}
\begin{tabular}{lccccc}
\hline
\hline
Aperture & Radius & $\tau_{\rm ff}$ & $\Sigma_{\rm mol}$ &  $\Sigma_{\rm SFR}$ & $\tau_{\rm dep}$ \\
 & (kpc) & (Myr)  & (M$_\odot$\,pc$^{-2}$) & (M$_\odot$\,Myr$^{-1}$\,pc$^{-2}$) & (Myr) \\
\hline
  1 &     6.05 &    18.9048 &     9.0602 &   0.002942 &    3080.04 \\
  2 &     5.59 &     7.5313 &    26.9317 &   0.011834 &    2275.83 \\
  3 &     5.60 &    12.0490 &    12.3808 &   0.005420 &    2284.34 \\
  4 &     6.07 &    10.9773 &    30.2223 &   0.005767 &    5240.82 \\
  5 &     6.91 &    14.4198 &    12.1143 &   0.003054 &    3966.48 \\
  6 &     8.00 &    12.3769 &     4.6325 &   0.004616 &    1003.55 \\
  7 &     5.28 &    12.2899 &     9.5922 &   0.004752 &    2018.68 \\
  8 &     3.98 &    10.9072 &    17.2317 &   0.007388 &    2332.38 \\
  9 &     3.02 &     8.2883 &    44.4875 &   0.012769 &    3483.95 \\
 10 &     2.80 &     8.5171 &    53.7099 &   0.012607 &    4260.41 \\
\hline
\end{tabular}
\label{table:hexagons}
\end{center}
\tablefoot{The columns correspond to the aperture galactocentric radius, ensemble-averaged free-fall time, molecular gas and SFR surface densities, and mean depletion time within hexagonal apertures of 1.5\,kpc diameter.}
\end{table*}

\subsection{Interferometric data}

As part of the IRAM S16BF project (P.I.\: A.~Schruba), IC\,342 was observed with 8~antennas in the D~configuration of NOEMA during 120\,hours: $111$\,hours in August 2016 and $9$\,hours in February 2017. Atmospheric turbulence leads to phase random variations that scatters flux. This effects can be quantitatively described with a millimeter ``seeing'' value. For our 3\,mm observations,  this seeing varied from $0.5\arcsec$ to $2.0\arcsec$, and the system temperature varied from ${\sim}150$ to ${\sim}400$\,K depending on the weather condition and source elevation. The field of view was covered with a mosaic of 941 pointings following a hexagonal pattern sampled at half the primary beam width size, that is, $21.45\arcsec$ at $115.271$\,GHz. The receiver bandwidth was correlated at $1.95$\,MHz channel spacing, i.e., about $5.1$\,km\,s$^{-1}$, by the WIDEX correlator. It was also correlated at $310$\,kHz channel spacing by the narrow band correlator, but only for 6 of the 8 antennas. Indeed, the NOEMA POLYFIX correlator was only commissioned end of 2017. In this letter, we only use the WIDEX data to maximize the sensitivity.

The distance covered by a visibility in the $uv$-plane during an integration should always be smaller than the distance associated to tolerable aliasing (see~\citealt{pety10} for more details). This can be written as $\tint \ll 6900/(\beam[fov]/\beam[syn]) \sim 46\,\emr{s}$, where $\beam[fov] \sim 600\arcsec$ is the field-of-view angular size, and $\beam[syn] \sim 4\arcsec$ the angular resolution (Eq.~(C.3) in this article). We also wish to loop over the maximum number of pointings in a given time, typically between two calibrations, so that the point spread function of the interferometer does not vary much from one pointing to the other. However, the slewing time between two pointings in stop-and-go mosaicking implies an overhead of about 10\,\emr{s}. We thus choose an integration time of 15\,\emr{s} as a compromise. Finally, we want each pointing to be sampled several times during one typical 10\,hour observing track. This implied to observe at most 160 pointings per track. We thus split the 941 pointings into 6 rectangular areas containing between 154 and 160 pointings each. Observing the full field-of-view thus required a minimum of 6 tracks and each rectangular area was observed during typically two tracks (or 20\,hours).

We used the quasars 0224$+$671 (typically $0.7$\,Jy) and J0359$+$600 (typically $0.6$\,Jy) as phase and amplitude calibrators, and observations of MWC\,349 and LKH$\alpha$\,101 as ``primary'' flux calibrators to set the Jy/K conversion for each antenna. We used the standard \texttt{GILDAS}/\linebreak[0]{}\texttt{CLIC}\footnote{See \url{http://www.iram.fr/IRAMFR/GILDAS} for more information about the \texttt{GILDAS} softwares.} pipeline to calibrate the data (calibration of the radio-frequency bandpass, phase as function of time, flux, and amplitude as a function of time). After flagging out about 20\% of data acquired during {unstable} weather, the calibrated $uv$ table contains the equivalent of 29\,hours on-source observations with 8~antennas.

In summary, the NOEMA observatory invested 120 hours with almost always
8~antennas. About one fifth of this time was observed during too {unstable}
summer weather to be useful. The standard observing efficiency (on-source
time divided by on-source plus overhead times) is 0.51 when using two flux
calibrators. Wide-field stop-and-go mosaicking divides the efficiency by
another factor 1.7, because of the time required to slew (acceleration and
deceleration time included) from one pointing to the other, leading to an
overall observing efficiency of 0.31. This implies about 29\,hours of
on-source observations with 8~antennas.

\mod{Since 2022, NOEMA is} 2.36 times faster to reach the same sensitivity thanks to its 12 available antennas. Moreover using the On-The-Fly observing mode instead of the stop-and-go mosaicking would enable to increase the observing efficiency by a factor 1.7. Assuming the same weather conditions, the same observation of IC\,342 would thus only require 26 hours of NOEMA telescope time.

\subsection{Short spacing information from \mbox{30-m}}

In order to provide the short spacings that were filtered out by the interferometer, we observed the galaxy with the IRAM \mbox{30-m} telescope during 43\,hours in July 2016 as part of project \mbox{083-16}. We used the On-The-Fly Position-Switching observing mode to image a $11\times 11\arcmin$ field of view. The FTS spectrometer delivered a channel spacing of 195\,kHz. The ON--OFF gain calibration was reprocessed using the \texttt{GILDAS}/\linebreak[0]{}\texttt{MRTCAL} software that was put in production in February 2017 at the telescope in order to use the possibility to compute one gain value every 20\,MHz instead of one every 4\,GHz. This provides a much more accurate gain calibration of the \chem{^{12}CO}{10} line that lies close to an atmospheric oxygen line. We went to the $T_\emr{mb}$ brightness scale using forward and beam efficiency values of 0.95 and 0.78. We then extracted a window of 300\,MHz around the $115.271$\,GHz CO frequency and we fit an order~2 polynomial baseline excluding a fixed velocity window of [$-100,+160$\,km\,s$^{-1}$]. Finally, we gridded the data using a convolution with a Gaussian kernel of size $7\arcsec$, delivering a cube of $22.5\arcsec$ angular resolution.

\subsection{Final \texorpdfstring{\chem{CO}{10}}{CO(1-0)} data cube}

The \texttt{GILDAS}/\linebreak[0]{}\texttt{MAPPING} software was used to create the short-spacing visibilities~\citep{pety10}. These pseudo-visibilities were merged with the interferometric (alone) observations in the $uv$ plane \mod{using the task \texttt{uvshort}}.  Each mosaic field was then imaged and a dirty mosaic was built. To speed up the long processing time of the full treatment of the inherent shift-variant response of interferometric wide-field observing mode, we divided the field-of-view in 16~tiles that overlap by two primary beamwidths. We deconvolved these tiles independently using the Högbom CLEAN algorithm, and we stitched the inner region together. In other words, each pixel was deconvolved taking into account the interferometric response up to at least one primary beamwidth. We regularized the found clean components during the Högbom deconvolution by a Gaussian of FWHM $4.00\arcsec \times 3.25\arcsec$ at a position angle of $91^\circ$. This synthesized beam is the minimum common resolution beam for the 941 individual pointings. We defined a broad mask \mod{for cleaning} from the deep IRAM \mbox{30-m} observations as follows: we smoothed the single dish cube to a resolution of $33\arcsec$ and we detected the signal using a dilated mask approach going to $2\sigma$ from all peaks above $5\sigma$. We deconvolved up to the point where the residual looks like noise inside the broad mask. A visual inspection of the result ensured that we deconvolved all significant signal. The resulting data cube has an rms of $0.11$\,K for a $5$\,km\,s$^{-1}$ wide channel.

\section{\texorpdfstring{\chem{CO}{10}}{CO(1-0)} derived products}
\label{sec:derived-products}

\subsection{Images}

We created the PHANGS standard list of derived products: a noise cube, smoothed versions of the line emission at fixed angular $(5, 15, 21\arcsec)$ and physical ($90, 150, 500, 1000, 1500$\,pc) resolution, a strict mask using a watershed method for noise estimation at each resolution, and a broad complete mask composed of the union of these masks, a set of moments (peak temperature, line integrated area, centroid velocity, line width, equivalent width) and their associated uncertainties. \mod{For the beam-by-beam analysis, we employ the CO intensity and line width derived consistently using the strict mask.} The creation of these products followed the PHANGS standard procedure described in \citet{LEROY21b}, except that we worked at $5, 10, 20$\,km\,s$^{-1}$ channel resolutions instead of $2.5$\,km\,s$^{-1}$ used for the PHANGS--ALMA data.

\subsection{Cloud catalog}
\label{app:clouds}

To identify molecular clouds, we ran the \texttt{CPROPS} algorithm \citep{ROSOLOWSKY06,ROSOLOWSKY21} on the 5\,km\,s$^{-1}$ data cube after convolving it to $\theta = 90$\,pc (FWHM) resolution. We used the implementation by \citet{ROSOLOWSKY21} and choose $\theta=90$\,pc to exactly match the physical resolution of their analysis. This allows a fair comparison to the measurements of GMC properties for the subset of PHANGS--ALMA galaxies that have resolution $\leq 90$\,pc by \citet{ROSOLOWSKY21} and A.~Hughes et al.\ (in preparation), though we do note that the PHANGS--ALMA cubes have velocity resolution $2.6$\,km\,s$^{-1}$, two times better than the NOEMA IC\,342 data. The IC\,342 $90$\,pc cube has noise of ${\sim}90$\,mK in \chem{CO}{10} per 5\,km\,s$^{-1}$ channels, and we compare to the ${\sim}80$\,mK \chem{CO}{21} in the $2.6$\,km\,s$^{-1}$ PHANGS--ALMA \chem{CO}{21} homogenized measurements from A.~Hughes et al.\ (in preparation). Accounting for $R_{21} \approx 0.65$ \citep[e.g.,][]{DENBROK21,LEROY21c} and the velocity difference, the equivalent \chem{CO}{10} noise at 5\,km\,s$^{-1}$ for PHANGS--ALMA is $87$\,mK, an excellent match to our IC\,342 data.

\citet{ROSOLOWSKY21} and \citet{ROSOLOWSKY06} describe the details of the \texttt{CPROPS} segmentation and cloud characterization algorithms; we follow \citet{ROSOLOWSKY21} exactly. Briefly, local maxima are identified and a mask is created in signal-to-noise space following the method of \citet{LEROY21a}. The method estimates a spatially and spectrally varying noise cube empirically from the data, $\sigma_T(x,y,v$), and identifies significant emission by finding all pixels with $T>4\sigma_T$ in two adjacent channels.  This mask is then dilated to all connected pixels with $T>2\sigma_T$ in two adjacent channels. The algorithm eliminates all local maxima in the map that are $<2\sigma_T$ above the \mod{merge level} containing at least one other maximum to suppress noise features. \mod{Following \citet{ROSOLOWSKY21}, given that cloud sizes are comparable to our synthesized beam, we impose no minimum separation between local maxima ($d_{\rm min}=0$ spatially and $v_{\rm min}=0$ spectrally). We also do not impose any thresholds ($s=0$) to consider if a pair of clouds is unique compared to merging them (\texttt{SIGDISCONT} parameter in the original CPROPS code). We require $N > 0.25 \Omega_{\rm bm}/\Omega_{\rm px}$ spaxels to be uniquely linked with a given local maximum ($\Omega_{\rm bm}$ and $\Omega_{\rm px}$ are the synthesized beam and pixel size, respectively).}

Then we segment the emission within the mask into individual clouds using a watershed algorithm  and measure their properties using moment methods, including radius, mass based on CO luminosity, and velocity dispersion. To estimate masses from the CO emission, $M_{\rm CO}$, we used a single ``Galactic'' conversion factor of 
$\alpha_{\rm CO} = 4.35\,M_\odot$\,pc$^{-2}$ (K\,km\,s$^{-1}$)$^{-1}$ \citep{BOLATTO13B}. \mod{We used the extrapolated and deconvolved cloud measurements to account for the finite sensitivity and limited resolution of the data. The extrapolation involves measuring cloud properties limited to progressively lower intensity thresholds, fitting this function, and evaluating it at an intensity threshold of 0 K. We deconvolve the NOEMA synthesized beam from the observations, assuming that the cloud and beam are both Gaussian. We derive radii based on the deconvolved major and minor moments following $R = \eta \sqrt{\sigma_{\rm maj,d} \sigma_{\rm min,d}}$ and adopt the same $\eta = 1.18$ factor as in \citet{ROSOLOWSKY21}, which is different from the $\eta = 1.91$ adopted by many previous studies. We compute the average surface density within the FWHM size of each cloud as $\Sigma_{\rm mol} = M_{\rm CO} / (2\pi R^2)$, based on (``2D'') the radius directly measured by CPROPS and extrapolated and deconvolved cloud mass (for a two dimensional Gaussian cloud model, 50\% of the mass is enclosed by the Gaussian FWHM). When computing volume densities (e.g., $\rho$ in $\tau_{\rm ff}$), we set the cloud radius to the $R_\mathrm{3D}$ prescription used in \citet{ROSOLOWSKY21}, where $R_\mathrm{3D} = R$ if $R \leq H / 2$, where $H$ is the molecular disk scaleheight (for simplicity, we adopt a constant $H = 100$\,pc ), and $R_\mathrm{3D} = \sqrt[3]{R^2 H / 2}$ if $R > H / 2$.}

\subsection{Definition of dynamical environments}
\label{app:environment}

\FigNIR{} %

{Following \citet{QUEREJETA21}, we used the near-IR morphology to distinguish four environments in IC\,342, the center, bar, 
spiral arm, and inter-arm regions (Fig.~\ref{fig:NIR}). 
The stellar bar was defined visually on a \textit{Spitzer} IRAC 3.6\,$\mu$m image downloaded from IRSA\footnote{https://irsa.ipac.caltech.edu} and correspond to an ellipse with semi-major axis $120''$, ellipticity 0.378, and $\mathrm{P.A.}=160^\circ$. The center environment is defined as an ellipse of semi-major axis $42.3''$, ellipticity 0.436, and $\mathrm{P.A.}=202^\circ$, and corresponds to an elongated bright region in IRAC 3.6\,$\mu$m, likely due to nonstellar emission in the structures surrounding the nucleus.} The arm region refers to spiral arms defined following the technique described in \citet{HERRERA-ENDOQUI15}. An unsharp-mask version of the IRAC 3.6\,$\mu$m image was used to identify bright regions along arms and a log-spiral function was fitted to these points. Using these log-spiral segments as ridge lines, the arms were assigned a width of 1000\,pc in the plane of the galaxy and the remainder of the non-arm, non-\mod{bar} galaxy was labeled as inter-arm. The spiral arms of IC\,342 do not show a very strong contrast at 3.6\,$\mu$m and we caution that the spiral regions here may not reflect arms with the same strength seen in other galaxies. Still, they provide a useful reference to study differences between different parts of the galaxy disk.

\end{appendix}

\end{document}